\documentclass[12pt]{article}

\usepackage{epsfig,latexsym,amsfonts,amsmath,amsthm,amssymb,amsbsy,multirow,slashed,wasysym,textcomp,subfigure,hyperref,wrapfig,datetime,comment,mathtools,cancel,cite,mathrsfs}
\usepackage[font={footnotesize,sl},bf]{caption}
\usepackage{braket}
\usepackage{calligra}

\def\ee{\end{equation}}
\def\bea{\begin{eqnarray}}
\def\eea{\end{eqnarray}}
\newcommand{\beq}{\begin{eqnarray}}
\newcommand{\eqq}{\end{eqnarray}}
\newcommand{\badat}{\begin{alignedat}}
\newcommand{\eadat}{\end{alignedat}}

\newcommand{\eal}[1]{\be \begin{aligned} #1 \end{aligned}\end{equation}} 
\newcommand{\eqn}[1]{\be #1 \end{equation}} 
\newcommand{\eqa}[1]{\bea  #1\end{eqnarray}}

\long\def\new#1\endnew{{\bf #1}}		
\long\def\del#1\enddel{}

\def\del{\partial}

\usepackage{xcolor}
\definecolor{oldmauve}{rgb}{0.4, 0.19, 0.28}
\definecolor{pansypurple}{rgb}{0.47, 0.09, 0.29}
\definecolor{burgundy}{rgb}{0.5, 0.0, 0.13}
\definecolor{carminepink}{rgb}{0.92, 0.3, 0.26}
\definecolor{blue(pigment)}{rgb}{0.2, 0.2, 0.6}
\definecolor{darkseagreen}{rgb}{0.56, 0.74, 0.56}
\definecolor{darkspringgreen}{rgb}{0.09, 0.45, 0.27}
\definecolor{ceruleanblue}{rgb}{0.16, 0.32, 0.75}

\newcommand{\p}{\partial}

\newcommand{\be}{\begin{eqnarray}}
\newcommand{\en}{\end{eqnarray}}

\def\bh{{\bar h}}
\def\bz{{\bar z}}
\def\bw{{\bar w}}
\def\bx{{\bar x}}
\def\by{{\bar y}}
\def\ba{{\bar a}}
\def\bb{{\bar b}}
\def\bc{{\bar c}}
\def\bd{{\bar d}}
\newcommand{\mY}{\mathcal{Y}}
\newcommand{\mbY}{\bar{\mathcal{Y}}}
\newcommand{\bpartial}{\bar\partial}
\author{Laura Donnay}
\numberwithin{equation}{section} % equation numbers follow sections

\begin{document}
\begin{titlepage}
  \thispagestyle{empty}
  
  \begin{center} 
  \vspace*{3cm}
{\LARGE\textbf{Logarithmic doublets in CCFT}}

\vskip1cm

   \centerline{Agnese Bissi$^{a,b,d}$\footnote{abissi@ictp.it}, Laura Donnay$^{c,d}$\footnote{ldonnay@sissa.it},  Beniamino Valsesia$^{c,d}$\footnote{bvalsesi@sissa.it}}
   
\vskip1cm

\it{$^a$ICTP, International Centre for Theoretical Physics, }\\
\it{Strada Costiera 11, 34151, Trieste, Italy}\\
\it{$^b$Department of Physics and Astronomy, Uppsala University,}\\
\it{Box 516, SE-751 20 Uppsala, Sweden}\\
\it{$^c$SISSA, Via Bonomea 265, 34136 Trieste, Italy}\\
\it{$^d$INFN, Sezione di Trieste, Via Valerio 2, 34127 Trieste, Italy}\\

\end{center}

\vskip1cm

\begin{abstract}
We investigate the presence of logarithmic CFT doublets in the soft sector of celestial CFT related with supertranslations. We show that the quantum operator associated with a  $\log u$ late-time behavior for the asymptotic gravitational shear forms a logarithmic CFT pair of conformal dimension $\Delta=1$ with an IR-regulated supertranslation Goldstone current. We discuss this result in connection with previous encounters of log CFT structures in the IR-finite part of celestial OPEs. \\

\end{abstract}
\end{titlepage}

\tableofcontents 
%\newpage
\section{Introduction}
Scattering amplitudes, the most basic observables of flat spacetimes, take the form of conformal correlators on the celestial sphere once rewritten in a basis that makes their $SL(2,\mathbb C)$ transformation properties manifest~\cite{deBoer:2003vf,Strominger:2017zoo,Pasterski:2016qvg,Pasterski:2017ylz}. The map amounts to trading the on-shell momentum of a bulk particle for the data of the point $(z,\bz)$ at which it pierces the celestial sphere and a conformal dimension $\Delta$. In that way, each incoming or outgoing particle is associated with an operator $\mathcal O_{\Delta}(z,\bz)$ on the celestial sphere which is thus identified as a `celestial CFT' (CCFT) operator (see \cite{McLoughlin:2022ljp,Pasterski:2021rjz,Raclariu:2021zjz,Donnay:2023mrd} for reviews on celestial holography).

Celestial CFT correlation functions, by their very definition, should be consistent with all known asymptotic symmetry constraints. The latter can be elegantly expressed in a CFT language as current algebras on the celestial sphere arising from insertions of celestial operators of a certain (sometimes negative) integer conformal dimension $\Delta$ (see e.g. \cite{Donnay:2018neh,Adamo:2019ipt,Guevara:2019ypd,Pate:2019mfs,Puhm:2019zbl,Fotopoulos:2019vac}). This `conformally soft' sector is governed, at leading order, by the Goldstone and memory modes associated with spontaneously broken supertranslation (for gravity) and large gauge (for gauge theory) symmetries. Both of these modes turn out to play a key role in the universal infrared (IR) properties of scattering amplitudes~\cite{Weinberg:1965nx}. Indeed, in gravity\footnote{In this work we focus on the gravitational case, but similar statements hold for gauge theory.}, a descendant of the memory mode gives rise to the supertranslation current, whose insertion into the $S$-matrix was famously shown to reproduce Weinberg's soft graviton theorem \cite{Strominger:2013jfa,He:2014laa}. On the other hand, the supertranslation Goldstone boson, denoted $C(z,\bar z)$, is responsible for the fact that amplitudes with an IR cutoff can be factorized into a hard and a soft piece, the latter containing all IR divergences \cite{Himwich:2020rro,Arkani-Hamed:2020gyp}. The deep lesson we have learned in the last years is that IR divergences are present in order to set to zero all amplitudes that violate BMS asymptotic symmetry conservation laws \cite{Kapec:2017tkm,Choi:2017bna,Choi:2017ylo}.
 IR divergences arising from virtual soft gravitons exchanges are accounted for by the supertranslation Goldstone $C(z,\bar z)$ current algebra. While the gravitational memory effect \cite{gravmem3,Strominger:2014pwa,Ashtekar:2014zsa} is an IR-safe observable, the Goldstone two-point function is logarithmic with an IR-divergent level proportional to the gravitational cusp anomalous dimension~\cite{Himwich:2020rro,Arkani-Hamed:2020gyp} (see also \cite{Nande:2017dba} for the QED analog). 

Studies of the supertranslation Goldstone mode have also played an important role in the following works. In \cite{Nguyen:2021ydb}, an effective action for supertranslation modes was derived from an analysis of the gravitational field equations and action principle near spatial infinity $i^0$. In particular, the supertranslation Goldstone mode was related there to the Goldstone associated with the breaking of `spi supertranslations' at $i^0$~\cite{Ashtekar:1978zz,Compere:2011ve}. In  \cite{Javadinezhad:2023mtp}, the authors solved the long-standing problem of deriving a covariant and supertranslation-invariant formula for the flux of angular momentum in asymptotically flat spacetimes. One of the key observations they have made is that any supertranslation-invariant flux is defined with respect to a particular point which defines the origin of the coordinate system and that one needs to define the lowest $l=0,1$ harmonics of the boundary gravitons at $u\to \pm \infty$  in order to completely define a covariant and supertranslation-invariant flux. In  \cite{Pasterski:2021dqe,Pasterski:2021fjn,Pano:2023slc}, the structure of conformal multiplets in CCFT was analyzed and, in the conformally soft sector, Goldstone modes correspond to the operators at the top of `celestial diamonds'. Because of their divergent logarithmic two-point function, the question was raised whether they should be excluded from the original CCFT spectrum or whether they could be somehow included in a logarithmic extension of the theory~\cite{Pano:2023slc}. Finally, a novel point of view on the CCFT soft sector was put forward in \cite{Kapec:2021eug,Kapec:2022axw,Kapec:2022hih}, where it was shown that the infinite-dimensional space of vacua in asymptotically flat spacetime can be thought of from a geometric perspective where soft insertions implement parallel transport about the infinite-dimensional vacuum manifold.

Logarithmic CFTs (log CFTs) are conformal field theories whose correlation functions might exhibit logarithmic singularities~\cite{Gurarie:1993xq}; see also \cite{Ferrara:1972jwi,Saleur:1991hk,Rozansky:1992td} for early works and \cite{Hogervorst:2016itc,Flohr2003BPlcft,Gaberdiel:2001tr,Creutzig:2013hma,Gurarie:2013tma} for reviews, including discussions of their relevance in the description of certain statistical models. They are characterized by the fact that they contain reducible but indecomposable representations of the conformal group, called logarithmic multiplets. A remarkable feature of log CFTs is the fact that the full conformal invariance of the theory is preserved despite the presence of a scale. This feature makes them a priori interesting to study in the context of celestial holography, given the persistence of a length scale in asymptotically flat spacetimes (in contrast to AdS spacetime where the presence of both the AdS radius and Newton constant allows to build dimensionless quantities).
In fact, encounters of logarithmic CFT patterns in the celestial CFT literature have recently appeared \cite{Bhardwaj:2022anh,Fiorucci:2023lpb,Bhardwaj:2024wld}. In \cite{Fiorucci:2023lpb}, a logarithmic partner of the celestial stress tensor \cite{Kapec:2016jld} was considered as a composite operator made out of the celestial stress tensor  and the superrotation Liouville field  \cite{Compere:2016jwb,Compere:2018ylh}. From the celestial OPE side, the presence of logarithms in the IR-finite parts of loop-corrected gluon OPEs was observed in \cite{Bhardwaj:2022anh,Bhardwaj:2024wld} (see also \cite{Krishna:2023ukw}). This has raised the possibility that certain sectors of celestial CFT could be organized by logarithmic CFT multiplets.

The goal of this paper is to investigate the presence of logarithmic CFT doublets in the soft sector of CCFT associated with supertranslations. It is structured as follows. In section \ref{sec:Celestial_CFT_operators}, we review the basic celestial CFT objects needed for our discussion, such as celestial operators as conformal primaries and soft (Goldstone and memory) modes associated to supertranslation symmetry. We also provide a new derivation of the Goldstone boson two-point function which was first inferred in \cite{Himwich:2020rro} from Weinberg's formula for the factorization of IR divergences from virtual gravitons. Section \ref{sec:LogCFT} provides a brief review of the definition of logarithmic CFT primaries and multiplets. We also introduce there a `log-shadow transform', a non-local transform which maps a log CFT doublet of weights $(h,\bar h)$ to another log CFT doublet of weights $(2-h,2-\bar h)$.  In section \ref{sec:log_doublet}, we show that the quantum operator associated with a $\log u$ behavior of the gravitational shear at late time induces a log CFT doublet structure in the supertranslation currents sector. We start by reviewing in section \ref{subsec:partner} the form of this quantum operator in terms of ladder operators, which was already pointed out in \cite{Campiglia:2019wxe,AtulBhatkar:2019vcb} in QED. We then show that this primary operator modifies the transformation property of the Goldstone supertranslation current to that of a logarithmic primary and that they form together a logarithmic pair of conformal weights $(\frac{3}{2},-\frac{1}{2})$. We then identify the log-primary in section \ref{subsec:two_point_functions} as an IR-regulated expression of the supertranslation Goldstone current and use this regulated mode to compute the two-point functions of the logarithmic pair. We end in section \ref{sec:discussion} with a discussion on our results.
In particular, we provide an interpretation of our two-point log CFT structure by drawing an analogy with the case of a free scalar in $2d$ CFT, where we show that a similar structure can be obtained when considering the operator extracting the divergent part of the correlator, which is related to the presence of the free scalar zero-mode. We also comment on the scenario in which logarithmic correlation functions can appear in Coulomb gas models when the Liouville puncture operator is included in the spectrum of a gravitationally dressed CFT.

\section{Celestial CFT operators}
\label{sec:Celestial_CFT_operators}
In this section, we present the generic set-up and give a brief introduction to the constructions of primary operators in celestial CFT.

\subsection{Massless fields in flat space}

We start by introducing the main set-up, following mostly the notation of \cite{Donnay:2022wvx}. We consider a massless bosonic spin-$s$ $(s=0,1,2,\dots)$ field $\phi_I^{(s)}$ in flat space  $\mathbb{R}^{1,3}$. The index $I=\mu_1\dots\mu_s$ is a collection of symmetrized indices, $(\mu_i=0,\dots, 3)$ in Cartesian coordinates $X^\mu=(t,\vec{x})$. At late times, the massless field admits the mode expansion
\begin{equation}\label{phis}
    \phi^{(s)}_I(X)=K_{(s)}\sum_{\alpha=\pm}\int\frac{d^3p}{(2\pi)^3 2p_0}\left[\epsilon_I^{*\alpha}(\vec q)a_\alpha^{(s)}(\vec p)e^{ip_\mu X^\mu}+\epsilon_I^{\alpha}(\vec q)a_{\alpha}^{(s)}(\vec p)^\dagger e^{-ip_\mu X^\mu}\right]\,
\end{equation}
with $K_{(s)}$ a constant and where the momentum is parametrized as (see e.g. \cite{He:2014laa})
\begin{gather}
    p^\mu=\omega q^\mu(w,\bw),\quad q^\mu=\frac{1}{\sqrt{2}}\left(1+w\bw,w+\bw,i(\bw-w),1-w\bw\right)\,
\end{gather}
and the polarization tensors $\epsilon^\pm_I(\vec q)=\epsilon^\pm_{\mu_1}(\vec q)\dots\epsilon^\pm_{\mu_s}(\vec q)$ with
\begin{gather}
    \epsilon^+_\mu(\vec q)=\partial_wq_\mu=\frac{1}{\sqrt{2}}(-\bw,1,-i,-\bw),\quad \epsilon^-_\mu(\vec q)=\epsilon^+_\mu(\vec q)^*=\partial_\bw q_\mu=\frac{1}{\sqrt{2}}(-w,1,-i,-w)\,.
\end{gather}
The ladder operators $a_\alpha^{(s)}(\vec p)$ satisfy the usual commutation relations, which in this basis read
\begin{equation}
    \left[a_\alpha^{(s)}(\omega,z,\bz),a_\beta^{(s)}(\xi,w,\bw)^\dagger\right]=\frac{16\pi^3}{\omega}\delta(\omega-\xi)\delta^{(2)}(z-w)\delta_{\alpha\beta}.
    \label{eq:as_commutation_relations}
\end{equation}
Using the transformation properties of the spin-$s$ field under Poincar\'e
\begin{equation}
    \phi'^{(s)}_{\mu_1\dots\mu_s}(X')={\Lambda_{\mu_1}}^{\nu^1}\dots{\Lambda_{\mu_s}}^{\nu^s} \phi^{(s)}_{\nu_1\dots\nu_s}(X),\quad X'^\mu={\Lambda_{\mu}}^{\nu}X^\nu+t^\mu\,,
\end{equation}
we can deduce the transformation of the ladder operators. In particular, the Lorentz group induces on the frequency and the holomorphic coordinates the following $SL(2,\mathbb{C})$ M\"obius action:
\begin{equation}
    w'=\frac{a w+b}{cw+d},\quad \omega'=\left|\frac{\partial w'}{\partial w}\right|^{-1}\omega,\quad q'^\mu=\left|\frac{\partial w'}{\partial w}\right|{\Lambda^\mu}_\nu q^\nu(w,\bw)\,,
    \label{eq::action_SL(2,C)_on_momentum}
\end{equation}
where ${\Lambda^\mu}_\nu$ is the Lorentz matrix related to the $SL(2,\mathbb{C})$ transformations. Ladder operators transform under Poincar\'e transformations as
\begin{equation}
    {a_\alpha^{(s)}}'(\omega',w',\bw')=\left(\frac{\partial w'}{\partial w}\right)^{-\frac{J}{2}}\left(\frac{\partial \bw'}{\partial \bw}\right)^{\frac{J}{2}}e^{-i\omega q^\mu(w,\bw){\Lambda_\mu}^\nu t_\nu}a_\alpha^{(s)}(\omega,w,\bw). 
    \label{eq::transformation_of_momentum_ladder}
\end{equation}

\subsection{Celestial primaries}
A conformal primary wavefunction is a solution of the linearized spin-$s$ equation of motion that transforms under $SL(2,\mathbb{C})$ as a 2$d$ conformal primary of weight $\Delta$ and spin $J$ and a 4$d$ spinor/tensor of spin $s$ \cite{Pasterski:2016qvg,Pasterski:2017kqt,Pasterski:2020pdk}:
\begin{equation}
    \phi_{\Delta,\alpha}^{(s)}\left({\Lambda^\mu}_\nu X^\nu,\frac{aw+b}{cw+d},\frac{\ba\bw+\bb}{\bc\bw+\bd}\right)=(cw+d)^{\Delta+J}(\bc\bw+\bd)^{\Delta-J}D_{s}(\Lambda)\phi_{\Delta,\alpha}^{(s)}\left(X^\mu,w,\bw\right)\,,
\end{equation}
where $\alpha=\pm$ is the helicity index, $J=\alpha s$, $D_s(\Lambda)$ is the $3+1\, d$ spin-$s$ representation of the Lorentz algebra and $\Lambda(M)$ is the Lorentz matrix related to the $SL(2,\mathbb{C})$ element:
\begin{equation}
    M=
    \begin{pmatrix}
        a & b \\
        c & d
    \end{pmatrix}
    \in SL(2,\mathbb{C}). 
\end{equation}
An operator on the celestial sphere is then constructed using the symplectic product \cite{Donnay:2020guq}
\begin{equation}
    \mathcal{O}^{(s),\pm}_{\Delta,\alpha}(w,\bw)=\left(\Phi^s(X),\phi_{\Delta^*,-\alpha}^{(s)}\left(X^\mu\mp i\epsilon,w,\bw\right)\right)_\Sigma\,,
\end{equation}
where the $\pm$ superscript indicates whether the operator is a creation or annihilation operator.\\
This formal definition is equivalent, up to a normalization constant (see e.g. \cite{Donnay:2022ijr} for details), to the Mellin transform of the usual momentum space operators $a^{(s)}_\alpha(\omega,z,\bz),\,a^{(s)\dagger}_\alpha(\omega,z,\bz)$
\begin{align}
\badat{2}
    \mathcal{O}^{(s),+}_{\Delta,\alpha}(z,\bz)\sim a_{\Delta,\alpha}^{(s)}(z,\bz)&=\int_0^{+\infty} d\omega\,\omega^{\Delta-1} a_{\alpha}^{(s)}(\omega,z,\bz),\\
    \mathcal{O}^{(s),-}_{\Delta,\alpha}(z,\bz)\sim a_{\Delta,\alpha}^{(s),\dagger}(z,\bz)&=\int_0^{+\infty} d\omega\,\omega^{\Delta-1}{a_{-\alpha}^{(s)}(\omega,z,\bz)}^\dagger
    \label{eq::Mellin_ladders}\,.
    \eadat
\end{align}
These operators act on the vacuum creating boost eigenstates, which form a basis on the space of square normalizable wavefunctions for $\Delta$ in the continuous principal series, $\Delta\in 1+i\lambda,\, \lambda\in\mathbb{R}$~\cite{deBoer:2003vf,Pasterski:2017kqt}. 
It is easy to prove that
the Mellin basis operators transform under Poincar\'e as
\begin{equation}
    {a_{\Delta,\alpha}^{(s)}}'(w',\bw')=\left(\frac{\partial w'}{\partial w}\right)^{-\frac{\Delta+J}{2}}\left(\frac{\partial \bw'}{\partial \bw}\right)^{-\frac{\Delta-J}{2}}e^{-iq^\mu(w,\bw){\Lambda_\mu}^\nu t_\nu\delta_\Delta}a_{\Delta,\alpha}^{(s)}(w,\bw)\,,
    \label{eq:Mellin_Basis_Transformation}
\end{equation}
where translations implement a shift in the conformal dimension $\delta_\Delta a_{\Delta,\alpha}^{(s)}:=a_{\Delta+1,\alpha}^{(s)}$ . Focusing on the $SL(2,\mathbb{C})$ part of Poincar\'e by fixing $t=0$, if we rewrite $w'=f(w),\,\bw'=\bar f(\bw)$,  we see that $a_{\Delta,\alpha}^{(s)}(w,\bw)$ transforms as a conformal primary field:
\begin{equation}
    a_{\Delta,\alpha}^{(s)'}(w',\bw')=\left(\del f\right)^{-h}\left(\bpartial\bar f\right)^{-\bh}a_{\Delta,\alpha}^{(s)}(w,\bw)
    \label{eq:Mellin_Basis_Conformal_Transformation}
\end{equation}
of weights
\begin{equation}
    h=\frac{\Delta+J}{2},\quad \bh=\frac{\Delta-J}{2}\,.
\end{equation}
The complex conjugate operator $a_{\Delta,-\alpha}^{(s),\dagger}(z,\bz)$ transforms in the same way with the swap $J\rightarrow-J$.
In what follows we will focus in particular on spin-two operators $a_{\Delta,\alpha}^{(2)}(z,\bz)$. 

\subsection{Soft gravitational phase space}
\label{sec:Phase_space_of_asymptotically_flat_spacetime}
The mode expansion for a graviton is read from \eqref{phis} with $s=2$ :
\begin{equation}\label{graviton}
    h_{\mu \nu}(X)=\kappa \sum_{\alpha=\pm}\int\frac{d^3p}{(2\pi)^3 2p_0}\left[\epsilon_{\mu\nu}^{*\alpha}(\vec q)a_\alpha(\vec p)e^{ip_\mu X^\mu}+\epsilon_{\mu\nu}^{\alpha}(\vec q)a_{\alpha}(\vec p)^\dagger e^{-ip_\mu X^\mu}\right]\,
\end{equation}
where $\kappa=\sqrt{32\pi G}$ and we dropped the spin superscript in $a, a^\dagger$. One can obtain the boundary field at future null infinity $\mathscr I^+$ by using retarded Bondi coordinates\footnote{We use the same coordinate conventions as in \cite{Himwich:2020rro}, where the Minkowski metric is given by $ds^2=-dudr+r^2dzd\bz$.} and taking the large $r$ limit. This leads to (see e.g. \cite{He:2014laa})
\begin{gather}
    C_{zz}=\frac{\kappa}{8i\pi^2}\int_0^{\infty}d\omega\left[a_+e^{-i\omega u}-a_-^\dagger e^{i\omega u}\right],\quad
    C_{\bz\bz}=\frac{\kappa}{8i\pi^2}\int_0^{\infty}d\omega\left[a_-e^{-i\omega u}-a_+^\dagger e^{i\omega u}\right] \,,\label{eq:Czz_ca_def}
\end{gather}
where the gravitational shear is defined as $C_{zz}(u,z,\bz)=\lim_{r\rightarrow\infty}r^{-1}h_{zz}(u,r,z,\bz)$ (and similarly for the $C_{\bz \bz}$ component).

The asymptotic shear is paired with the news tensor $N_{zz}=\partial_u C_{zz}$ through the Ashtekar-Streubel symplectic structure \cite{Ashtekar:1978zz,Ashtekar:1981bq,Ashtekar:1981sf}. Together, they describe  the `hard'  part (as opposed to the `soft' piece described below) of the gravitational phase space and satisfy the following commutation relations\footnote{In $N_{\bz\bz}(u,z,\bz)$ and $C_{ww}(u',w,\bw)$ we dropped the dependence on $z$ and $\bz$ to make the notation more compact.}: 
\begin{equation}
    \left[N_{\bz\bz}(u),C_{ww}(u')\right]=-16\pi i G\delta(u-u')\delta^{(2)} (z-w)\,.
\end{equation}

\quad Under the action of the extended BMS group (see \cite{Donnay:2023mrd} for a recent review on BMS symmetries in the context of celestial holography), they transform as~\cite{Barnich:2010eb}
\begin{align}  \delta_{(\mathcal{T},\mY,\mbY)}C_{zz}&=\left(\mY\partial+\mbY\bpartial+\frac{3}{2}\partial\mY-\frac{1}{2}\bpartial\mbY\right)C_{zz}+\left(\mathcal{T}+\frac{\partial\mY+\bpartial\mbY}{2}\right)N_{zz}-2\partial^2\mathcal{T}-u\partial^3\mY\\
    \delta_{(\mathcal{T},\mY,\mbY)}N_{zz}&=\left(\mY\partial+\mbY\bpartial+2\partial\mY\right)N_{zz}+\left(\mathcal{T}+\frac{\partial\mY+\bpartial\mbY}{2}\right)\partial_u N_{zz}-\partial^3\mY\,,
\end{align}
where $\mathcal T(z,\bz)$ and $\mathcal Y(z),\mbY(\bz)$ denote the supertranslation and superrotation\footnote{We allow for conformal Killing vector fields with possible isolated poles on the celestial sphere~\cite{Barnich:2010eb}.} generators, respectively.\\

\noindent \textbf{Goldstone boson}\\
\vspace{-0.5cm}

\noindent The supertranslation Goldstone mode, denoted $C(z,\bz)$, is related to the infinite vacua degeneracy of asymptotically flat spacetimes \cite{He:2014laa,Strominger:2017zoo}. It was shown to play a key role in the context of (conformal) dressing leading to IR-finite amplitudes \cite{Himwich:2020rro,Arkani-Hamed:2020gyp}, as we will briefly review below.
It owes its name to the fact that it transforms as a pure shift under the action of an infinitesimal supertranslation, $\delta_{\mathcal T}C(z,\bz)=\mathcal T(z,\bz)$ while it transforms under superrotations as a primary of weights $(-\frac{1}{2},-\frac{1}{2})$ 
\cite{Compere:2016jwb,Himwich:2020rro}:
\begin{equation}
    \delta_{(\mY,\mbY)} C(z,\bz)=\left(\mY\partial+\mbY\bpartial-\frac{1}{2}\partial\mY-\frac{1}{2}\bpartial\mbY\right)C(z,\bz)\,.
\end{equation}

An expression of the supertranslation Goldstone boson in terms of creation and annihilation operators can be found
in \cite{Arkani-Hamed:2020gyp}: 
\begin{equation}
\badat{2}
    C(z,\bz)
    &=\frac{i\kappa}{8\pi^2}\int_0^\infty d\omega\int \frac{d^2y}{2\pi}\,\Big[\,\frac{z-y}{\bz-\by}\left(a_+(\omega,y,\by)-a_-^\dagger(\omega,y,\by)\right)\\
    &\quad \quad \quad \quad \quad \quad  \quad \quad  \quad \quad   +\frac{\bz-\bx}{z-x}\left(a_-(\omega,y,\by)-a_+^\dagger(\omega,y,\by)\right)\Big]\\
    &=\frac{i\kappa}{8\pi^2}\int_0^\infty d\omega\int \frac{d^2y}{2\pi}\left[\frac{z-y}{\bz-\by}\,\delta a_+ +\frac{\bz-\by}{z-y} \,\delta a_-\right]\,,
    \label{eq:C_as_expression}
\eadat
\end{equation}
with $\delta a_\pm:=a_\pm-a_\mp^\dagger$.
Using the commutation relations \eqref{eq:as_commutation_relations} one can check that $C(z,\bz)$ satisfies the commutation relation\footnote{$\Theta(u)=\text{sign}(u)$ denotes the sign function with the property: $\partial_u\Theta(u)=2\delta(u)$.}:
\begin{align}
    \left[C(z,\bz),C_{ww}(u)\right]&=-8\pi i G\frac{\bz-\bw}{z-w}\Theta(u)\,.
    \label{eq:C_commutation_relations}
\end{align}

We also recall that from $C(z,\bz)$ descends an additional primary field known as the Goldstone supertranslation current (see e.g. \cite{Himwich:2020rro,Pasterski:2021dqe}):
\begin{equation}\label{partial_z^2}
\mathcal C_{zz}(z,\bz)=\partial_z^2 C(z,\bz)\,.
\end{equation}
It is thus a $(\frac{3}{2},-\frac{1}{2})$ primary descendant and, using $\partial_z \frac{1}{\bz-\by}=2\pi\delta^{(2)}(z-y)$, one can write \cite{Arkani-Hamed:2020gyp}
\begin{equation}\label{Czz}
\badat{2}
\mathcal C_{zz}(z,\bz)
    =\frac{i\kappa}{8\pi^2}\int_0^\infty d\omega\,\, &\Big[a_+(\omega,z,\bz)-a_-^\dagger(\omega,z,\bz)\\
    &+\int \frac{d^2y}{\pi}\frac{\bz-\by}{(z-y)^3}\left(a_-(\omega,z,\bz)-a_+^\dagger(\omega,z,\bz)\right)\Big]\,,
\eadat
\end{equation}
and satisfies $\partial_\bz^2 \mathcal C_{zz}=\partial_z^2 \mathcal C_{\bz\bz}$ and $\mathcal C_{\bz\bz}^\dagger=\mathcal C_{zz}$. Notice that expression \eqref{Czz} is the combination of a $\Delta=1$, $J=2$ mode and the shadow transform\footnote{The shadow transform is defined in eq. \eqref{eq::primary_shadow}.} of a $\Delta=1$, $J=-2$ primary. Also, the relation \eqref{partial_z^2} implies that for a given expression of $\mathcal C_{zz}$ there is a global translation ambiguity in the Goldstone mode $C(z,\bz) \to C(z,\bz) + t$ since a global (Poincar\'e) translation $t$ is such that $\p_z^2 t=0$. \\

\noindent \textbf{Memory modes}\\
\vspace{-0.5cm}

\noindent The (leading) soft graviton operator $\mathcal{N}_{zz}^{(0)}$ (and similarly its complex conjugate $\mathcal{N}_{\bz\bz}^{(0)}$) is defined as the following zero mode of the news tensor~\cite{He:2014laa,Kapec:2014opa}
\begin{align}
    \mathcal{N}_{zz}^{(0)}&=\int_{-\infty}^{\infty}du N_{zz}(u)=C_{zz}|_{\mathscr I^+_+}-C_{zz}|_{\mathscr I^+_-}\,.
    \label{eq:leading_soft_mode_Nzz_def}
\end{align}
Since it can be expressed as a difference of the $u\to \pm \infty$ values of the shear, it is sometimes referred to as the `memory mode'. 
It transforms as a $(\frac{3}{2},-\frac{1}{2})$ primary under the extended BMS symmetries
\begin{equation}
    \begin{split}
       &\delta_{(\mathcal T, \mY)} \mathcal{N}^{(0)}_{zz} = \left(\mathcal Y \partial + \mbY \bar \partial + \frac{3}{2} \partial \mY - \frac{1}{2} \bar \partial \mbY \right) \mathcal{N}^{(0)}_{zz} \,,\\
    \end{split} 
    \label{eq:soft_mode_variations}
\end{equation}
while $\mathcal{N}_{\bz\bz}^{(0)}$ is a $(-\frac{1}{2},\frac{3}{2})$ primary.
Its commutation relations with the shear are
\begin{align}
    \left[\mathcal{N}^{(0)}_{zz},C_{\bw \bw}(u)\right]&=\left[\mathcal{N}^{(0)}_{\bz\bz},C_{ww}(u)\right]=-16\pi i G\delta^{(2)}(z-w)\,.
    \label{eq:N0_commutation_relations}
\end{align}
From \eqref{eq:leading_soft_mode_Nzz_def} we can also read off the usual expressions \cite{He:2014laa,Donnay:2022wvx}
\begin{equation}\label{N0modes}
    \mathcal{N}^{(0)}_{zz}=-\frac{\kappa}{4\pi}\lim_{\omega\rightarrow0}\omega\left[a_+(\omega)+a_-^\dagger(\omega)\right],\quad \mathcal{N}^{(0)}_{\bz\bz}=-\frac{\kappa}{4\pi}\lim_{\omega\rightarrow0}\omega\left[a_-(\omega)+a_+^\dagger(\omega)\right]\,.
\end{equation}
Using the commutation relations \eqref{eq:as_commutation_relations} one can prove that these modes satisfy \eqref{eq:N0_commutation_relations}. 
The two above operators can be combined to obtain the memory operators~\cite{Pasterski:2021dqe}
\begin{equation}
\badat{2}
    \mathcal{N}_{zz}
    &=\frac{1}{2\pi}\left[\mathcal{N}_{zz}^{(0)}+\int \frac{d^2w}{\pi}\frac{\bz-\bw}{(z-w)^3}\,\mathcal{N}_{\bw\bw}^{(0)}\right]\\
    \mathcal{N}_{\bz\bz}
    &=\frac{1}{2\pi}\left[\mathcal{N}_{\bz\bz}^{(0)}+\int \frac{d^2w}{\pi}\frac{z-w}{(\bz-\bw)^3}\,\mathcal{N}_{ww}^{(0)}\right],
\eadat
\end{equation}
of conformal weights $\left(\frac{3}{2},-\frac{1}{2}\right)$ and $\left(-\frac{1}{2},\frac{3}{2}\right)$, respectively. 
They also satisfy
\begin{equation}
    \left[\mathcal{N}_{zz},C(w,w)\right]=-\frac{4G}{\pi}\frac{\bz-\bw}{z-w},\quad \left[\mathcal{N}_{\bz\bz},C(w,w)\right]=-\frac{4G}{\pi}\frac{z-w}{\bz-\bw}\,.
\end{equation}
Alternatively, the memory operators can be obtained as level-two primary descendants of the memory scalar mode, namely 
\begin{equation}
    \mathcal{N}_{zz}=\partial_z^2N(z,\bz),\quad \mathcal{N}_{\bz\bz}=\partial_\bz^2N(z,\bz)\,,
\end{equation}
with $N(z,\bz)$ the $\left(-\frac{1}{2},-\frac{1}{2}\right)$ primary defined as~\cite{Pasterski:2021dqe}
\begin{equation}
    N(z,\bz)=\frac{1}{2\pi}\left[\int \frac{d^2w}{2\pi}\frac{z-w}{\bz-\bw}\mathcal{N}_{zz}^{(0)}+\int \frac{d^2w}{2\pi}\frac{\bz-\bw}{z-w}\,\mathcal{N}_{\bw\bw}^{(0)}\right]\,.
    \label{eq:N_generalized}
\end{equation}

The first descendant of the leading soft graviton operator is referred to as the supertranslation current $P_z$
\begin{equation}\label{soft_operators}
    P_z=\frac{2}{\kappa^2}\bpartial \mathcal{N}^{(0)}_{zz},
\end{equation}
whose insertion in the $S$-matrix was shown to give rise to Weinberg's soft graviton theorem ~\cite{Strominger:2013jfa,He:2014laa}.
Using the definition of the modes in terms of creation and annihilation operators, one easily computes the two-point functions involtving $C$ and $P_z$.
From a CFT point of view, these are equivalent to the OPEs~\cite{Himwich:2020rro}:
\begin{equation}\label{PC OPE}
    P_zC(w,\bw)\sim-\frac{i}{z-w},\quad P_zP_w\sim 0\,.
\end{equation}
The first OPE is precisely compatible with that of a current with its Goldstone mode, while the second one states that the current is invariant under supertranslations.\\

\noindent \textbf{Revisiting the Goldstone two-point function}\\
\vspace{-0.5cm}
\label{sec:Revisiting_Goldstone_twopf}

\noindent We now turn to the computation of the Goldstone boson two-point function $\langle CC\rangle$ explicitly from the modes $\eqref{eq:C_as_expression}$.
Let us first notice that, from \eqref{eq:as_commutation_relations}, we read that
\begin{equation}
\badat{2}
    \langle \delta a_\pm(z,\bz,\omega) \delta a_\pm(w,\bw,\xi)\rangle&=0\\
    \langle \delta a_\pm(z,\bz,\omega) \delta a_\mp(w,\bw,\xi)\rangle&=-\frac{16\pi^3}{\omega}\delta(\omega-\xi)\delta^{(2)}(z-w)\,,
\eadat
\label{eq:delta_a_two_point}
\end{equation}
so that the two-point function reduces to
\begin{align}
   & \langle C(z,\bz)C(w,\bw)\rangle\\\nonumber
    &=\frac{-\kappa^2}{4(2\pi)^4}\int_0^\infty d\omega\int_0^\infty d\xi\int \frac{d^2x}{2\pi}\frac{d^2y}{2\pi}\left[\frac{z-x}{\bz-\bx}\frac{\bw-\by}{w-y}\langle\delta a_+(x)\delta a_-(y)\rangle+\frac{\bz-\bx}{z-x}\frac{w-y}{\bw-\by}\langle\delta a_-(x)\delta a_+(y)\rangle\right]\\\nonumber
    &=\frac{\kappa^2}{2(2\pi)^3}\int_0^\infty \frac{d\omega}{\omega}\int d^2x\left[\frac{z-x}{\bz-\bx}\frac{\bw-\bx}{w-x}+\frac{\bz-\bx}{z-x}\frac{w-x}{\bw-\bx}\right]\nonumber\,.
\label{eq:CC_unregularized}
\end{align}
The integrals over $\omega$ and $(x,\bx)$ diverge so they need to be regularized.  
We are interested in the IR behaviour and will neglect UV divergences. We will use dimensional regularization (dim-reg) with $d=2+2\epsilon$ to treat the IR divergences in the $(x,\bx)$ integral. Notice that this also cures the IR divergence in $\omega$ as in $d$-dimension the two-point function \eqref{eq:delta_a_two_point} becomes
\begin{equation}
    \langle \delta a_\pm(z,\bz,\omega) \delta a_\mp(w,\bw,\xi)\rangle=-\frac{16\pi^3}{\omega^{d-1}}\delta(\omega-\xi)\delta^{(d)}(\vec{z}-\vec{w})\,.
\end{equation}
We will then focus on the following regulated expression:
\begin{equation}
    C_\epsilon(z,\bz)=\frac{i\kappa}{8\pi^2}\int_0^\infty d\omega\omega^{2\epsilon}\, \int \frac{d^{2+2\epsilon}w}{2\pi} \left[\frac{z-w}{\bz-\bw}\,\delta a_+(w,\bw,\omega) +\frac{\bz-\bw}{z-w}\,\delta a_-(w,\bw,\omega)\right]\,.
    \label{eq:Regulated_C_mode}
\end{equation}
Notice that in the limit $\epsilon\rightarrow0$ we recover the expression for the Goldstone mode $C$.  

The regulated two-point function is thus
\begin{align}
    \langle C_\epsilon(z,\bz)C_\epsilon(w,\bw)\rangle
    &=\frac{\kappa^2\mu_0^{2\epsilon}}{2(2\pi)^3}\int_0^\Lambda d\omega \omega^{-1+2\epsilon}\int d^{2+2\epsilon}x\left[\frac{z-x}{\bz-\bx}\frac{\bw-\bx}{w-x}+\frac{\bz-\bx}{z-x}\frac{w-x}{\bw-\bx}\right]\,,
\end{align}
where $\mu_0$ is the mass scale introduced in dim-reg to preserve the mass dimension of the two-point function, and $\Lambda$ is the UV cutoff. The $\omega$ integral reduces to
\begin{equation}
    \int_0^\Lambda d\omega \omega^{-1+2\epsilon}=\frac{\Lambda^{2\epsilon}}{2\epsilon}=\frac{1}{2\epsilon}+\log\Lambda\sim\frac{1}{2\epsilon}\,,
\end{equation}
neglecting as announced the UV divergence. Hence,
\begin{equation}\label{cc1}
    \langle C_\epsilon(z,\bz)C_\epsilon(w,\bw)\rangle
    =\frac{\kappa^2}{4(2\pi)^3}\frac{\mu_0^{2\epsilon}}{\epsilon}\int d^{2+2\epsilon}x\left[\frac{z-x}{\bz-\bx}\frac{\bw-\bx}{w-x}+\frac{\bz-\bx}{z-x}\frac{w-x}{\bw-\bx}\right]\,.
\end{equation}
The integral
\begin{equation}
    I_\epsilon:=\mu_0^{2\epsilon}\int d^{2+2\epsilon}x\,\frac{z-x}{\bz-\bx}\frac{\bw-\bx}{w-x}\,
    \label{eq:I_Epsilon}
\end{equation}
can then be computed using standard QFT techniques and gives the following result (see appendix \ref{sec:Regulated_integral_computation}): 
\begin{equation}
    I_\epsilon=\frac{6\pi^{2+\epsilon}(1+\epsilon)\Gamma(2+\epsilon)^2}{\Gamma(1+\epsilon)\Gamma(4+2\epsilon)\sin(\pi\epsilon)}|z-w|^{2+2\epsilon}\mu_0^{2\epsilon}.
\end{equation}
This lead to the following expression for the two-point function:
\begin{align}
\badat{2}
    \langle C_\epsilon(z,\bz)C_\epsilon(w,\bw)\rangle &=\frac{\kappa^2}{16\pi^3}\frac{1}{\epsilon}I_\epsilon\\
    &=\frac{3\kappa^2}{8\pi^3\epsilon}\frac{\pi^{2+\epsilon}(1+\epsilon)\Gamma(2+\epsilon)^2}{\Gamma(1+\epsilon)\Gamma(4+2\epsilon)\sin(\pi\epsilon)}|z-w|^{2+2\epsilon}\mu_0^{2\epsilon}\,.  
    \eadat
    \label{eq:dim_reg_CC}
\end{align}
The form of this two-point function is that of a primary field of weights $\left(-\frac{1+\epsilon}{2},-\frac{1+\epsilon}{2}\right)$, namely we see that the dimensional regularization induces a shift in the conformal dimension of $C$.
We can then expand \eqref{eq:dim_reg_CC} for  small  $\epsilon$ and get:
\begin{equation}
\badat{2}
    \langle C_\epsilon(z,\bz)C_\epsilon(w,\bw)\rangle
    &=\frac{\kappa^2}{16\pi^2}|z-w|^2\frac{1}{\epsilon}\left(\frac{1}{\epsilon}-\frac{2}{3}+\gamma_E+\log\pi+\log|z-w|^2\mu_0^2\right)+\dots\\
    &=\frac{\kappa^2}{16\pi^2}|z-w|^2\left(\frac{1}{\epsilon^2}+\frac{1}{\epsilon}\log|z-w|^2\mu^2\right)+\dots\,,
\label{eq:CC_regulated_two_point_function}
\eadat
\end{equation}
where $\mu^2=\mu_0^2\pi e^{\gamma_E-\frac{2}{3}}$ and the dots denote finite terms in limit $\epsilon\rightarrow0$.

\quad A few comments are now in order: the second term in the expression coincides with the value computed in \cite{Himwich:2020rro}\footnote{The extra $\mu^2$ factor compared to \cite{Himwich:2020rro} can be reabsorbed in the UV regulator.}.  Notice that the prefactor $\frac{\kappa^2}{16\pi^2 \epsilon }=\frac{1}{\epsilon}\frac{2G}{\pi}$ matches with the gravitational cusp anomalous dimension which appears in the level of the Goldstone current (see also \cite{Nande:2017dba,Miller:2012an} for related works). In \cite{Himwich:2020rro}, the supertranslation Goldstone two-point function was originally derived from Weinberg's factorization formula \cite{Weinberg:1965nx} in the following way: soft factorization states that a scattering amplitude $\mathcal A$ with an IR cutoff can be expressed as a product of a soft $\mathcal S$-matrix (which includes all IR divergences) and a hard piece (which is IR finite), 
\begin{equation}
\mathcal A=\mathcal A_{\text{soft}}\mathcal A_{\text{hard}}\,.
\end{equation}
The two-point function of the Goldstone supertranslation current $C$ is such that the correlators of vertex operators $\mathcal W_k=e^{i\eta_k \omega_k C(z_k,\bar z_k)}$ exactly account for the soft part, namely
\begin{equation}
\mathcal A_{\text{soft}}=\langle \mathcal W_1 \dots  \mathcal W_n\rangle \,.
\end{equation}

We see that \eqref{eq:CC_regulated_two_point_function} also contains an extra leading divergent piece of the form $\frac{1}{\epsilon^2}|z-w|^2$. It is important to note that this term does not spoil the above soft factorization. Actually, such a divergence is also present in Weinberg's soft (virtual particle) exponential for massless states, but this term eventually drops out from the final expression due of total momentum conservation\footnote{We are grateful to Sruthi Narayanan for clarifying this point.}. We expect that the extra $\epsilon^{-2}$ term in our derivation is scheme dependant and can be reabsorbed in the next $\epsilon^{-1}$ term.   
Finally, while the Goldstone two-point function is clearly divergent, it is important to note that the gravitational memory effect (see e.g. \cite{gravmem3,Strominger:2014pwa,Ashtekar:2014zsa}) is determined by the $P_z C$ OPE \eqref{PC OPE}, and is therefore an IR safe observable \cite{Himwich:2020rro}.

\section{Log CFT}
\label{sec:LogCFT}
In this section, we aim to give a short and self-contained summary of properties of logarithmic CFTs; see for instance \cite{Flohr:2001zs,Hogervorst:2016itc}. Log CFTs get their name from the particular behaviour of their correlation functions which contain logarithms of the spacetime coordinates. This peculiarity is due to the fact that, for these theories, the representation of the conformal group is not irreducible but falls into a reducible but indecomposable representation. This means that the dilatation operator $D$ cannot be completely diagonalized but it can be at most put into a Jordan block form such that the primary states fall into rank $r$ multiplets ($a=1,\dots,r$),
\begin{equation}
    D\ket{\mathcal{O}_a}=-i\boldsymbol{\Delta}_a^b\ket{\mathcal{O}_b},\quad \boldsymbol{\Delta}=
    \begin{pmatrix}
        \Delta & 1      & 0      & \cdots & 0      \\
        0      & \Delta & 1      & \cdots & 0      \\
        \vdots & \vdots & \vdots & \ddots & 0      \\
        0      & 0      & 0      & \Delta & 1      \\
        0      & 0      & 0      & 0      & \Delta \\
    \end{pmatrix}
    .
\end{equation}
It follows then that all of these theories are non-unitary. Indeed, reflection positivity would require the two-point function $\langle \mathcal O_a(x) \mathcal O_a(-x)\rangle$ to be positive, for all $x$ and (Hermitian) fields $\mathcal O_a$. However, for log CFTs, this is not possible, unless all multiplets have rank $r=1$. This follows from the fact that the dilatation operator is not Hermitian, thus in particular time translation on the cylinder is not implemented by a unitary operator.\\
In what follows, we will focus on two-dimensional log CFTs and give a brief exposition on the field representation of the conformal group in log CFTs.

\subsection{Log primaries and log CFT doublets}
In logarithmic CFTs, states are organized into logarithmic multiplets of rank $r\geq 1$. 
A logarithmic doublet $\mathcal O_a=(\Psi,\Phi)$ of weights $(h,\bar h)$ is composed of a \textit{primary operator} $\Phi$ and its \textit{logarithmic partner} $\Psi$ which transform under the global conformal group as \cite{Gurarie:1993xq}:
\begin{equation} 
\badat{2}
    &\Psi'\left(f(z),\bar f(\bar z)\right)=(\partial f)^{-h}(\bar \partial \bar f)^{-\bar h}\left(\Psi(z,\bz)-\log|\partial f|\Phi(z,\bz)\right),\\
    &\Phi'\left(f(z),\bar f(\bar z)\right)=(\partial f)^{-h}(\bar \partial \bar f)^{-\bar h}\Phi(z,\bz)
    \label{eq:log_doublet_tranformations}
\eadat
\end{equation}
where $f(z), \bar f(\bar z)$ are elements of the $SL(2,\mathbb{C})$ group. If this transformation extends to the full Virasoro algebra, we refer to these fields as \textit{Virasoro primary} and \textit{log Virasoro primary}. 
Using \eqref{eq:log_doublet_tranformations}, we can also find the general equation for the infinitesimal transformations of the logarithmic partner. Writing $f(z)=z-\mY(z),\ \bar f(\bz)=\bz-\mbY(\bz)$, we get
\begin{equation} 
\badat{2}
    \Psi'(z-\mY,\bz-\mbY)
    &=(1-\partial\mY)^{-h}(1-\bpartial\mbY)^{-\bh}\left[\Psi(z,\bz)-\frac{1}{2}\log\left((1-\partial\mY)(1-\bpartial\mbY)\right)\Phi(z,\bz)\right]\\
    &=(1+h\partial\mY)(1+\bh\bpartial\mbY)\left[\Psi(z,\bz)+\frac{\partial\mY+\bpartial\mbY}{2}\Phi(z,\bz)\right]\,.
\eadat
\end{equation}
Then if we consider $z\rightarrow z+\mY,\ \bz\rightarrow \bz+\mbY$ the field transformation becomes
\begin{align}   
    \Psi'(z,\bz)=(1+h\partial\mY)(1+\bh\bpartial\mbY)\left[(1+\mY\partial+\mbY\bpartial)\Psi+\frac{\partial\mY+\bpartial\mbY}{2}\Phi\right]\,,
\end{align}
so that the transformation properties of the log doublet can be written as
\begin{equation}
\badat{2}\label{log_doublet_def}
&\delta\Psi(z,\bz)=\left(\mY\partial+\mbY\bpartial+h\partial\mY+\bh\bpartial\mbY\right)\Psi(z,\bz)+\frac{1}{2}(\partial\mY+\bpartial\mbY)\Phi(z,\bz)\\
   & \delta\Phi(z,\bz)=\left(\mY\partial+\mbY\bpartial+h\partial\mY+\bh\bpartial\mbY\right)\Phi(z,\bz)\,.
\eadat\end{equation}
These relations constrain the $T\Psi$ OPE with the stress tensor to be
\begin{equation}
    T(z)\Psi(w,\bw)=\frac{1}{(z-w)^2}\left(h\Psi(w,\bw)+\frac{1}{2}\Phi(w,\bw)\right)+\frac{\partial\Psi(w,\bw)}{z-w}+\dots
    \label{eq:log_primary_T_OPE}
\end{equation}

If we identify $\Psi=\mathcal{O}_1,\,\Phi=\mathcal{O}_2$, then the two-point function of a logarithmic scalar doublet can always be written in the form \cite{Gurarie:1993xq,Gurarie:2013tma}
\begin{equation}
    \langle\mathcal{O}_a(z_1,\bar z_1)\mathcal{O}_b(z_2,\bar z_2)\rangle=\frac{1}{z_{12}^{2h}\,\bar z_{12}^{2\bar h}}
    \begin{pmatrix}
    \tilde{k}_{\mathcal O}-k_{\mathcal O}\log|z_{12}|^2 & k_{\mathcal O}\\
    k_{\mathcal O} & 0
    \end{pmatrix}_{ab}\,,
    \label{eq:log_doublet_two_point_function}
\end{equation}
with $k_{\mathcal O}\neq0$ and $\tilde{k}_{\mathcal O}$ some constant that cannot be fixed by conformal invariance. We remark that $\Psi$ is not uniquely specified because we may add to it any multiple of $\Phi$, $\Psi\rightarrow\Psi+k\Phi$, without affecting its defining properties, but changing the constant $\tilde{k}_{\mathcal O}$ in \eqref{eq:log_doublet_two_point_function} while $k_{\mathcal O}$ will remain invariant. Because of this, $\tilde{k}_{\mathcal O}$ may be tuned to any desired value, so it is not expected to be physical~\cite{Creutzig:2013hma}. The constant $k_{\mathcal O}$, on the other hand, is expected to be physically meaningful.
Another remarkable feature of a log CFT which can be observed from \eqref{eq:log_doublet_two_point_function} is that the two-point function of the two primary fields vanishes, $\langle \Phi \Phi \rangle=0$.

\subsection{Scale invariance}

The presence of logarithms would seem to signal that the $n$-point function is now scale dependent, as the quantity inside logarithms has to be dimensionless. However, this scale is physically irrelevant due to the Ward identities induced by global conformal invariance, as we review below (see \cite{Hogervorst:2016itc}).

Let us start exploring the idea in the context of standard irreducible CFTs.
Suppose that we analyze the system at a fixed reference scale $\mu$, the basis of primary operators at this scale frame will then be defined as $\mathcal{O}_{i,\mu}(z)$. Of course, the $n$-point function should not depend on this scale so we can fix a Callan-Symanzik equation to be such that
\begin{equation}
    \mu\frac{d}{d\mu}\langle\mathcal{O}_{i_1,\mu}(z_1)\dots\mathcal{O}_{i_n,\mu}(z_n)\rangle=\sum_{k=1}^n\langle\mathcal{O}_{i_1,\mu}(z_1)\dots\mu\frac{d}{d\mu}\mathcal{O}_{i_k,\mu}(z)\dots\mathcal{O}_{i_n,\mu}(z_n)\rangle=0.
\end{equation}
The total derivative with respect to the scale can be explicitly written as
\begin{equation}
    \frac{d}{d\mu}\mathcal{O}_{i,\mu}(z)=\lim_{\delta\mu\rightarrow0}\frac{\mu}{\delta\mu}\left[\mathcal{O}_{i,\mu+\delta\mu}\left(z+\frac{\delta\mu}{\mu}z\right)-\mathcal{O}_{i,\mu}(z)\right]\,,
\end{equation}
where we have taken into account the coordinate scale variation. The expression in the brackets is then the standard primary field variation under a rescaling and we can use its transformation properties to rewrite
\begin{equation}
    \mu\frac{d}{d\mu}\mathcal{O}_{i,\mu}(z)=\lim_{\delta\mu\rightarrow0}\frac{\mu}{\delta\mu}\left[z\frac{\delta\mu}{\mu}\partial\mathcal{O}_{i,\mu}\left(z\right)+h_i\frac{\delta\mu}{\mu}\mathcal{O}_{i,\mu}(z)\right]=z\partial\mathcal{O}_{i,\mu}(z)+h_i\mathcal{O}_{i,\mu}(z).
\end{equation}
Thus, it follows that the Callan-Symanzik equation turns exactly into the dilatation Ward identity,
\begin{equation}
     \mu\frac{d}{d\mu}\langle\mathcal{O}_{i_1,\mu}(z_1)\dots\mathcal{O}_{i_n,\mu}(z_n)\rangle=\sum_{k=1}^n(z_k\partial_{z_k}+h_k)\langle\mathcal{O}_{i_1,\mu}(z_1)\dots\mathcal{O}_{i_n,\mu}(z_n)\rangle=0\,,
\label{eq:irrelevance_of_the_scale}
\end{equation}
which is trivially satisfied due to conformal invariance.

Even in a standard CFT, it is of course natural to introduce a reference scale, and all the operators will then be defined at that scale. However, due to conformal invariance, the theory will not depend on the value assigned at the specific chosen scale, namely the field basis is dependent on the original choice of scale but the theory is not.
The exact same thing happens in a logarithmic CFT, even if the construction is more subtle due to the presence of logarithms. In this case, the different transformation properties of log primaries lead to a Callan-Symanzik equation of the form~\cite{Hogervorst:2016itc}
\begin{gather}\label{Callan}
     \mu\frac{d}{d\mu}\langle\mathcal{O}_{i_1,l_1;\mu}(z_1)\dots\mathcal{O}_{i_n,l_n;\mu}(z_n)\rangle=
     \sum_{k=1}^n\langle\mathcal{O}_{i_1,l_1;\mu}(z_1)\dots\left(\delta^{j_k}_{l_k}z_k\partial_{z_k}+\boldsymbol{h}^{j_k}_{l_k}\right)\mathcal{O}_{j_k,\mu}(z_k)\dots\mathcal{O}_{i_n,\mu}(z_n)\rangle=0
\end{gather}
where
\begin{equation}
     \boldsymbol{h}=
     \begin{pmatrix}
         h      & 1      & 0      & \dots  & 0      & 0      \\
         0      & h      & 1      & \dots  & 0      & 0      \\
         0      & 0      & h      & \dots  & 0      & 0      \\
         \vdots & \vdots & \vdots & \ddots & \vdots & \vdots \\
         0      & 0      & 0      & \dots  & h      & 1      \\
         0      & 0      & 0      & \dots  & 0      & h
     \end{pmatrix}
     .
\end{equation}
The RHS of \eqref{Callan} turns out to coincide with the dilatation Ward identity for a logarithmic primary field which must be satisfied in a logarithmic CFT, suggesting again that any $n$-point function will be independent on the chosen reference scale.
We can see an explicit realization of this fact by considering the two-point function of logarithmic field in a doublet $(\tilde{\mathcal{O}}_{h,\mu},\mathcal{O}_{h,\mu})$ at a chosen scale $\mu$:
\begin{equation}
    \langle\tilde{\mathcal{O}}_{h,\mu}(z)\tilde{\mathcal{O}}_{h,\mu}(0)\rangle=-\frac{\kappa}{(\mu z)^{2h}}\log(\mu z).
    \label{transfscale}
\end{equation}
This can be rewritten using \eqref{eq:log_doublet_tranformations} as
\begin{equation}
    \langle\mu^{h}\left(\tilde{\mathcal{O}}_{h,\mu}+\log\mu\mathcal{O}_{h,\mu}\right)(z)\mu^{h}\left(\tilde{\mathcal{O}}_{h,\mu}+\log\mu\mathcal{O}_h\right)(0)\rangle=-\frac{\kappa}{z^{2h}}\log(z).
    \label{transfscale1}
\end{equation}
By equating \eqref{transfscale} and \eqref{transfscale1} we can see that, while the form of the logarithmic primary field depends on the specific scale $\mu$, we can fix $\mu$ to an arbitrary value and the theory will still be invariant under the conformal group.

\subsection{Log-shadow transform}

The shadow transform of a conformal primary $\Phi$ of weights $(h,\bar h)$ is defined as (see e.g.~\cite{Osborn:2012vt})
\begin{equation}
\widetilde \Phi(z,\bz)=K_{h,\bh}\int d^2w \frac{\Phi (w,\bw)}{(z-w)^{2-2h}(\bz-\bw)^{2-2\bar h}},  
\label{eq::primary_shadow}
\end{equation} 
with the shadow normalization\footnote{The normalization is chosen here such that $\widetilde{\widetilde{\Phi}}=\Phi$.} $K_{h,\bh}=\frac{\Gamma(2-2\bar h)}{\pi\Gamma(2h-1)}$. The shadow operator $\widetilde \Phi$ is still a conformal primary but now of weights $(1-h,1-\bar h)$.
Given a logarithmic primary $\Psi$ of weights $(h,\bar h)$, we define its `logarithmic shadow transform' as
\begin{align}
    &S_{\log}\left[\Psi\right](z,\bz)=-K_{h,\bh}\int d^2w \frac{\Psi(w,\bw)+\mathrm{log}|z-w|^2\Phi(w,\bw)}{(z-w)^{2-2h}(\bz-\bw)^{2-2\bh}}\,,
\end{align}
and $S_{\log}\left[\Phi\right](z,\bz)$ being the shadow defined as \eqref{eq::primary_shadow}.
We prove in appendix \ref{sec:more_on_the_log_shadow} that $S_{\log}\left[\Psi\right](z,\bz)$ transforms as a logarithmic primary field under $SL(2,\mathbb{C})$ with $\widetilde \Phi(z,\bz)$ a primary field. We also check that it squares to
\begin{equation}
    S_{\log}\left[ S_{\log}\left[\Psi\right]\right](z,\bz)=(-1)^{-4\bar h}\left[\Psi(z,\bz)+\left(\frac{1}{1-2h}+\frac{1}{1-2\bh}-2\pi i\right)\Phi(z,\bz)\right]\,,
\end{equation}
which allows us to define the inverse log-shadow $S_{\log}^{-1}$ of a logarithmic doublet as
\begin{equation}
\begin{alignedat}{2}
    &S_{\log}^{-1}\left[\Psi\right](z,\bz)=(-1)^{4\bar h}\left[S_{\log}\left[\Psi\right](z,\bz)-\left(\frac{1}{1-2h}+\frac{1}{1-2\bh}-2\pi i\right)\widetilde\Phi(z,\bz)\right]\\
    &S_{\log}^{-1}\left[\Phi\right](z,\bz)=(-1)^{4\bar h}\widetilde\Phi(z,\bar z)\,.
\end{alignedat}
\end{equation}

\section{Supertranslation log doublet}
\label{sec:log_doublet}
In this section, we turn to the study of log CFT structures in the context of the soft sector of CCFT associated to supertranslation. We show that the presence of a $\log u$ piece in the radiative data at future null infinity $\mathscr I^+$ gives rise to a log CFT doublet associated with an IR-regulated Goldstone supertranslation current. 

\subsection{Supertranslation Goldstone and its partner}
\label{subsec:partner}
Let us consider the following combinations of soft Fourier modes
\begin{equation}\label{new}
    \mathcal{B}_{zz}(z,\bz)=\frac{i\kappa}{8\pi^2}\lim_{\omega\rightarrow 0}\omega(a_+(\omega,z,\bz)-a_-^\dagger(\omega,z,\bz)),\quad \mathcal{B}_{\bz\bz}(z,\bz)=\frac{i\kappa}{8\pi^2}\lim_{\omega\rightarrow 0}\omega(a_-(\omega,z,\bz)-a_+^\dagger(\omega,z,\bz))\,.
\end{equation}
As we will see below, from the gravitational phase space point of view, these modes correspond to the presence of a $\log u$ piece in the radiative data at the corners of $\mathscr I^+$,
\begin{equation}
    C_{zz}(u,z,\bz) \stackrel{u\to +\infty}{\sim}2\mathcal{B}_{zz}(z,\bz)\,\log u+\dots
    \label{Bmode}
\end{equation}
This mode and its analog for QED has been studied in \cite{Campiglia:2019wxe,AtulBhatkar:2019vcb}\footnote{The QED analog of $\mathcal{B}_{zz}$ was denoted $\overset{\ln}{A_{z}}$ in \cite{Campiglia:2019wxe}.}. It was argued there that, while absent in the classical theory, such a $\log u$ mode  turns out to be non-vanishing in the quantum theory.
Most interestingly, such a term for QED was shown to be associated with the $\tau^{-1}$ decay of Coulombic modes as they approach timelike infinity, and the quantization of \eqref{new} leads to an asymptotic charge which reproduces the quantum part of Sahoo-Sen's logarithmic\footnote{See also \cite{Bern:2014oka,Ciafaloni:2018uwe,Saha:2019tub,Krishna:2023fxg,Alessio:2024wmz,Alessio:2024onn} for other works on logarithmic soft theorems.} corrections \cite{Sahoo:2018lxl} to the subleading soft photon theorem \cite{Campiglia:2019wxe,AtulBhatkar:2019vcb}. It was also argued in \cite{AtulBhatkar:2019vcb} that including the term \eqref{Bmode} does not amount to introducing a new independent mode in the quantum system, as $\mathcal{B}_{zz}$ is fixed in terms of the classical free data.

The relation between soft operators of the form \eqref{new} and the $\log u$ fall-off at $u \to +\infty$ was first presented in \cite{Campiglia:2019wxe}; we give below an alternative proof of this relationship (which holds for gravity but is also readily adapted for the QED case). The aim is thus to show that $\mathcal{B}_{zz}$ can be expressed in terms of creation and annihilation operators as in \eqref{new}, under the assumption that the fall-offs of the shear \eqref{eq:Czz_ca_def} (and thus the news tensor) are given by
\begin{equation} \label{eqasympt}
\badat{2}
    C_{zz}(u,z,\bar z)& \stackrel{u\to +\infty}{=}uN_{zz}^{vac}(z)+C_{zz}^{+}(z,\bz)+2\mathcal{B}_{zz}(z,\bz)\log u+o(1/u),\\
    N_{zz}(u,z,\bar z)& \stackrel{u\to +\infty}{=} N_{zz}^{vac}(z)+\frac{2}{u} \mathcal{B}_{zz}(z,\bz)+o(1/u^2),\\
    C_{zz}(u,z,\bar z)& \stackrel{u\to -\infty}{=}uN_{zz}^{vac}(z)+C_{zz}^{-}(z,\bz)+o(1/u),\\
    N_{zz}(u,z,\bar z)& \stackrel{u\to -\infty}{=} N_{zz}^{vac}(z)+o(1/u^2)\,,
\eadat
\end{equation}
where the Goldstone currents can be written as $C_{zz}^\pm=-2\mathcal D^2_z C^\pm$ with $\mathcal D_z$ being the superrotation-covariant derivative \cite{Campiglia:2021bap,Donnay:2021wrk} and $C^\pm$ the supertranslation Goldstone boson at $\mathscr I^\pm$. Notice that we included for generality in the above expressions the `vacuum news' $N_{zz}^{vac}$ considered in \cite{Donnay:2021wrk,Campiglia:2021bap,Campiglia:2020qvc,Compere:2020lrt}, but what follows does not depend on whether this term is present or not. 
The first step is to analyze the Fourier transform of the shear \eqref{eq:Czz_ca_def}, which is given by
\begin{equation}
\badat{2}
    \tilde{f}(\omega,z,\bz)
    &=\int_{-\infty}^{+\infty}du e^{i\omega u-|u|\epsilon}C_{zz}(u,z,\bz)\\
    %&=\frac{\kappa}{8i\pi^2}\int_0^\infty d\xi\int_{-\infty}^{+\infty}du\, \left[a_+ e^{i(\omega-\xi) u-|u|\epsilon}-a_-^\dagger e^{i(\omega+\xi) u-|u|\epsilon}\right]=\\
    &=\frac{\kappa}{4i\pi}\int_0^\infty d\xi\,\left[a_+ \delta(\omega-\xi)-a_-^\dagger \delta(\omega+\xi)\right]=\frac{\kappa}{4i\pi}\left[a_+(\omega)\theta(\omega)-a_{-}^{\dagger}(-\omega)\theta(-\omega)\right]\,,
    \label{C_zz_fourier_transf}
\eadat
\end{equation}
where $\epsilon$ is a regulator meant to be small. 
By inspecting the relation above, we can see that showing \eqref{new} is equivalent to prove the following equality 
\begin{equation} \label{Bwithf}
    \mathcal{B}_{zz}=-\frac{1}{2\pi}\lim_{\omega\rightarrow0^+}\omega\left(\tilde{f}(\omega)+\tilde{f}(-\omega)\right)\,.
\end{equation}
In the remaining part of this section, we are going to show how to prove \eqref{Bwithf}.
%The crucial point is to notice that the behaviour of the shear and the news tensors have different asymptotics.
We start by analysing $\tilde{f}(\omega)$ which can be rewritten using integration by parts as
\begin{equation}
\badat{3}
    \tilde{f}(\omega)
    =&\int_{-\infty}^{0}du\, e^{i(\omega-i\epsilon)u}C_{zz}+ \int_{0}^{+\infty}du\, e^{i(\omega+i\epsilon)u}C_{zz}\\
    =&\frac{1}{i\omega-\epsilon}\int_{0}^{\infty}du\, \frac{d}{du}\left[e^{i(\omega+i\epsilon)u}C_{zz}\right]-\frac{1}{i\omega-\epsilon}\int_{0}^{\infty}du\, e^{i(\omega+i\epsilon)u}N_{zz}\\
    &+\frac{1}{i\omega+\epsilon}\int_{-\infty}^0 du\, \frac{d}{du}\left[e^{i(\omega-i\epsilon)u}C_{zz}\right]-\frac{1}{i\omega+\epsilon}\int_{-\infty}^0 du\, e^{i(\omega-i\epsilon)u}N_{zz}\\
     =&2\pi\delta(\omega)C_{zz}(0)+\frac{i}{\omega-i\epsilon}\int_{-\infty}^0 du\, e^{i(\omega-i\epsilon)u}N_{zz}+\frac{i}{\omega+i\epsilon}\int_0^{+\infty} du\, e^{i(\omega+i\epsilon)u}N_{zz}\,,
\eadat
\end{equation}
where we used the shear fall-off conditions to see that the total derivative will always be suppressed as $u$ goes to infinity.
We can then study the behavior of 
\begin{equation}
\badat{2}
    \omega\left[\tilde{f}(\omega)+\tilde{f}(-\omega)\right]
    &=4\pi\omega\delta(\omega)C_{zz}(0)\\
    &+\frac{i\omega}{\omega-i\epsilon}\int_{-\infty}^0 du\, e^{i(\omega-i\epsilon)u}N_{zz}+\frac{i\omega}{\omega+i\epsilon}\int_0^{+\infty} du\, e^{i(\omega+i\epsilon)u}N_{zz}\\
    &-\frac{i\omega}{\omega+i\epsilon}\int_{-\infty}^0 du\, e^{-i(\omega+i\epsilon)u}N_{zz}-\frac{i\omega}{\omega-i\epsilon}\int_0^{+\infty} du\, e^{-i(\omega-i\epsilon)u}N_{zz}
    \label{f+f}
\eadat
\end{equation}
in the limit $\omega\rightarrow 0$. 
We would like to firstly analyse the following terms in \eqref{f+f} 
\begin{equation} \label{twoterms}
    \frac{i\omega}{\omega-i\epsilon}\int_{-\infty}^0 du\, e^{i(\omega-i\epsilon)u}N_{zz}-\frac{i\omega}{\omega+i\epsilon}\int_{-\infty}^0 du\, e^{-i(\omega+i\epsilon)u}N_{zz}.
\end{equation}
In order to do so, we split the integrals into two regions $(-\infty,-1/\Lambda)\cup (-1/\Lambda,0]$, where $\Lambda$ is a small but positive finite number. We can use \eqref{eqasympt} and the asymptotic behavior of the news tensor to evaluate the integrals, thus allowing \eqref{twoterms} to be written in the region $(-\infty,-1/\Lambda)$ as
\begin{equation}
\badat{2}
    &\frac{i\omega N_{zz}^{vac}}{\omega-i\epsilon}\int_{-\infty}^{-1/\Lambda} du\, e^{i(\omega-i\epsilon)u}-\frac{i\omega N_{zz}^{vac}}{\omega+i\epsilon}\int_{-\infty}^{-1/\Lambda} du\, e^{-i(\omega+i\epsilon)u}\\
    &=N_{zz}^{vac}\omega\left[\frac{e^{-\frac{\epsilon+i\omega}{\Lambda}}}{(\omega-i\epsilon)^2}-\frac{e^{-\frac{\epsilon-i\omega}{\Lambda}}}{(\omega+i\epsilon)^2}\right].
\eadat
\end{equation}
If we expand for small $|\omega\pm i\epsilon|$, we get
\begin{equation}
\badat{2}
    &N_{zz}^{vac}\omega\left[\frac{1}{(\omega-i\epsilon)^2}-\frac{1}{(\omega+i\epsilon)^2}+\frac{i\Lambda}{\omega-i\epsilon}-\frac{i\Lambda}{\omega+i\epsilon}\right]+o(\omega)\\
    &=\, 2\pi N_{zz}^{vac}\omega\left[-i\delta'(\omega)+\Lambda\delta(\omega)\right]+o(\omega)\,.
    \label{-inf_inf_lambda}
\eadat
\end{equation}
This is vanishing in the limit of small frequencies so we do not get contributions in the interval $(-\infty,-1/\Lambda)$.
We then can focus on the region $(-1/\Lambda,0]$. If we consider the news tensor to be regular inside $(-1/\Lambda,0]$ then we can expand the exponential for $|\omega\pm i\epsilon|\ll\Lambda$ and get
\begin{equation}
\badat{2}
    &\frac{i\omega}{\omega-i\epsilon}\int_{-1/\Lambda}^0 du\, e^{i(\omega-i\epsilon)u}N_{zz}-\frac{i\omega}{\omega+i\epsilon}\int_{-1/\Lambda}^0 du\, e^{-i(\omega+i\epsilon)u}N_{zz}\\
    =&-\left(\frac{i\omega}{\omega+i\epsilon}-\frac{i\omega}{\omega-i\epsilon}\right)\int_{-1/\Lambda}^0 du\, N_{zz}-2\omega\int_{-1/\Lambda}^0 du\,u N_{zz}+o(\omega^2)\\
    =&\, -2\pi\omega\delta(\omega)\int_{-1/\Lambda}^0 du\, N_{zz}-2\omega\int_{-1/\Lambda}^0 du\,u N_{zz}+o(\omega^2)\,,
    \label{-inf_lambda_0}
\eadat
\end{equation}
which is also vanishing in the limit $\omega\rightarrow0^+$. We conclude then that \eqref{f+f} receives no contribution from the integral between $(-\infty,0]$ for small frequencies.\\
We now analyse the other terms in \eqref{f+f} in the same way
\begin{equation}
    \frac{i\omega}{\omega+i\epsilon}\int_0^{+\infty} du\, e^{i(\omega+i\epsilon)u}N_{zz}-\frac{i\omega}{\omega-i\epsilon}\int_0^{+\infty} du\, e^{-i(\omega-i\epsilon)u}N_{zz}.
\end{equation}
Splitting again the integration region into $[0,1/\Lambda)\cup(1/\Lambda,+\infty)$, we can perform the same steps. The main difference is that in the regime $(1/\Lambda,+\infty)$ the news tensor has a different asymptotic behaviour, which makes
\begin{align}
    &\frac{i\omega}{\omega+i\epsilon}\int_{1/\Lambda}^{+\infty} du\, e^{i(\omega+i\epsilon)u}N_{zz}-\frac{i\omega}{\omega-i\epsilon}\int_{1/\Lambda}^{+\infty} du\, e^{-i(\omega-i\epsilon)u}N_{zz}\\
    =&\,\frac{i\omega}{\omega+i\epsilon}\int_{1/\Lambda}^{+\infty} du\, e^{i(\omega+i\epsilon)u}\left(N_{zz}^{vac}+\frac{2}{u}  \mathcal{B}_{zz} \right)-\frac{i\omega}{\omega-i\epsilon}\int_{1/\Lambda}^{+\infty} du\, e^{-i(\omega-i\epsilon)u}\left(N_{zz}^{vac}+\frac{2}{u}  \mathcal{B}_{zz} \right)\nonumber\\
    =&\frac{i\omega}{\omega+i\epsilon}\left[\frac{e^{-\frac{\epsilon-i\omega}{\Lambda}}}{\epsilon-i\omega}N_{zz}^{vac}+  2\mathcal{B}_{zz} \Gamma\left(0,\frac{\epsilon-i\omega}{\Lambda}\right)\right]-\frac{i\omega}{\omega-i\epsilon}\left[\frac{e^{-\frac{\epsilon+i\omega}{\Lambda}}}{\epsilon+i\omega}N_{zz}^{vac}+  2\mathcal{B}_{zz} \Gamma\left(0,\frac{\epsilon+i\omega}{\Lambda}\right)\right]\nonumber
\end{align}
where $\Gamma(x,y)$ is the incomplete gamma function. Expanding the expression above for small $|\omega\pm i\epsilon|$ we obtain
\begin{equation}
\badat{1}
    &-2\pi\omega N_{zz}^{vac}\left(i\delta'(\omega)+\Lambda\delta(\omega)\right)\\
    &+2\omega  \mathcal{B}_{zz}\left\{\frac{i}{\omega+i\epsilon}(-\gamma_E-\log \Lambda(\epsilon-i\omega))-\frac{i}{\omega-i\epsilon}(-\gamma_E-\log \Lambda(\epsilon+i\omega))\right\}+o(\omega)\\
    =&-2\pi \omega N_{zz}^{vac}\left(i\delta'(\omega)+\Lambda\delta(\omega)\right)-(\gamma_E+\log\Lambda)4\pi  \mathcal{B}_{zz} \omega\delta(\omega)\\
    &-2i  \omega  \mathcal{B}_{zz} \left[\frac{\log(\epsilon-i\omega)}{\omega+i\epsilon}-\frac{\log(\epsilon+i\omega)}{\omega+i\epsilon}\right]+o(\omega).
\eadat
\end{equation}
By inspecting the regions with positive and negative frequencies, it is possible to show that 
\begin{equation}
    -i\left[\frac{\log(\epsilon-i\omega)}{\omega+i\epsilon}-\frac{\log(\epsilon+i\omega)}{\omega+i\epsilon}\right]=-\pi P\frac{1}{|\omega|}+2\pi P\frac{1}{|\omega|}\theta(-|\omega|)-2\pi\delta(\omega)P\log|\omega|
\end{equation}
which implies that
\begin{equation}
\badat{1}
    &\frac{i\omega}{\omega+i\epsilon}\int_{1/\Lambda}^{+\infty} du\, e^{i(\omega+i\epsilon)u}N_{zz}-\frac{i\omega}{\omega-i\epsilon}\int_{1/\Lambda}^{+\infty} du\, e^{-i(\omega-i\epsilon)u}N_{zz}\\
    &=-2\pi \omega N_{zz}^{vac}\left(i\delta'(\omega)+\Lambda\delta(\omega)\right)-\gamma_E 4\pi  \mathcal{B}_{zz} \omega\delta(\omega)\\
    &-2\mathcal{B}_{zz}\pi\omega P\frac{1}{|\omega|}+4\pi  \mathcal{B}_{zz}  \omega P\frac{1}{|\omega|}\theta(-|\omega|)-4\pi \mathcal{B}_{zz}\omega\delta(\omega)P\log|\Lambda\omega|\,.
     \label{+inf_lambda_inf}
\eadat
\end{equation}
The only non vanishing term in the limit $\omega\rightarrow0^+$ is the first one in the third line\footnote{$\lim_{\omega\rightarrow0}\theta(-|\omega|)=0$}, namely
\begin{equation}
    \lim_{\omega\rightarrow0}\left[\frac{i\omega}{\omega+i\epsilon}\int_{1/\Lambda}^{+\infty} du\, e^{i(\omega+i\epsilon)u}N_{zz}-\frac{i\omega}{\omega-i\epsilon}\int_{1/\Lambda}^{+\infty} du\, e^{-i(\omega-i\epsilon)u}N_{zz}\right]=-\pi\mathcal{B}_{zz}\,.
\end{equation}
As in the other case, no contributions are coming from the region $[0,1/\Lambda)$.\\
Combining all the pieces, we therefore obtain that
\begin{equation}
    \mathcal{B}_{zz}=-\frac{1}{2\pi}\lim_{\omega\rightarrow 0^+}\omega\left(\tilde{f}(\omega)+\tilde{f}(-\omega)\right)=\frac{i\kappa}{8\pi^2}\lim_{\omega\rightarrow 0}\omega\left(a_+(\omega)-a_-^\dagger(\omega)\right)\,.
\end{equation}

\subsection{Transformation properties}
\label{transformation_properties}
Under an infinitesimal conformal transformation, the shear transforms as a $(\frac{3}{2},-\frac{1}{2})$ `quasi-conformal Carrollian primary'\footnote{We use the terminology introduced in \cite{Donnay:2022wvx} for tensors living at $\mathscr I$; see also \cite{Ciambelli:2018wre,Ciambelli:2019lap,Freidel:2021qpz}.} 
\begin{equation}
    \delta C_{zz}=\left(\mY\partial+\mbY\bpartial+\frac{3}{2}\partial\mY-\frac{1}{2}\bpartial\mbY\right)C_{zz}+\frac{u}{2}(\partial\mY+\partial\mbY)N_{zz}-u\partial^3\mY\,.
\end{equation}
This corresponds to a finite transformation of the form
\begin{equation}
    C_{zz}'(u',z',\bz')=(\partial f)^{-\frac{3}{2}}(\bpartial \bar{f})^{\frac{1}{2}} C(u,z,\bz)+(\partial f)^{\frac{1}{2}}(\bpartial \bar{f})^{\frac{1}{2}}\,u S(f,z)\,,
    \label{C_finite_tranformation}
\end{equation}
where $z'=f(z), \bz'=\bar{f}(\bz), u'=|\partial f|u$ and $S(f,z)=\frac{f'''(z)}{f'(z)}-\frac{3}{2}\left(\frac{f''(z)}{f'(z)}\right)^2$ denotes the Schwarzian derivative. 

From this property, we can then deduce the transformations of the asymptotic components of $C_{zz}$ in the limit $u\rightarrow+\infty$ under a superrotation.
Using \eqref{eqasympt}, we have
\begin{equation}
\badat{2}
    &C_{zz}'(u',z',\bz')
   % &=u'{N^{vac}_{zz}}'+{C^+_{zz}}'+2\log u'\,   \mathcal{B}'_{zz} +\widetilde{C}'_{zz}(u',z',\bz')\\
    =|\partial f| u{N^{vac}_{zz}}'+{C^+_{zz}}'+2\log u\,   \mathcal{B}'_{zz} +2\log|\partial f|\,  \mathcal{B}'_{zz} +\widetilde{C}'_{zz}(u',z',\bz')\\
    &C_{zz}(u,z,\bz)
    =uN^{vac}_{zz}+C^+_{zz}+2\log u\,   \mathcal{B}_{zz} +\widetilde{C}_{zz}(u,z,\bz)\,,
\eadat
\end{equation}
where $\widetilde{C}_{zz}$ denotes the $o(u^{-1})$ terms.
Using \eqref{C_finite_tranformation} we can match the two expressions:
\begin{equation}
\badat{2}
     &|\partial f| u{N^{vac}_{zz}}'+{C^+_{zz}}'+2\log u\,   \mathcal{B}'_{zz} +2\log|\partial f|\, \mathcal{B}'_{zz}+\widetilde{C}'_{zz}(u',z',\bz')\\
    =&(\partial f)^{-\frac{3}{2}}(\bpartial \bar{f})^{\frac{1}{2}}\left[uN^{vac}_{zz}+C^+_{zz}+2\log u\,   \mathcal{B}_{zz} +\widetilde{C}_{zz}(u,z,\bz)\right]+(\partial f)^{\frac{1}{2}}(\bpartial \bar{f})^{\frac{1}{2}}S(f,z)u\,,
\eadat
\end{equation}
implying the following finite transformation laws
\begin{equation}
\badat{2}
   & {N_{zz}^{vac}}'=(\partial f)^{-2}N^{vac}_{zz}+S(f,z)\\
   & {C_{zz}^+}'=(\partial f)^{-\frac{3}{2}}(\bpartial \bar{f})^{\frac{1}{2}}\left(C_{zz}^+-2\log|\partial f|  \mathcal{B}_{zz} \right)\\
   & {  \mathcal{B}_{zz} }'=(\partial f)^{-\frac{3}{2}}(\bpartial \bar{f})^{\frac{1}{2}}  \mathcal{B}_{zz} \\
   & \widetilde{C}'_{zz}(u',z',\bz')=(\partial f)^{-\frac{3}{2}}(\bpartial \bar{f})^{\frac{1}{2}}\widetilde{C}_{zz}(u,z,\bz).
\eadat
\end{equation}
These can be written in infinitesimal form as
\begin{equation}
    \badat{2}
   & \delta{N_{zz}^{vac}}=(\mY\partial+2\partial\mY)N_{zz}^{vac}-\partial^3\mY\\
   & \delta{C_{zz}^+}=\left(\mY\partial+\mbY\bpartial+\frac{3}{2}\partial\mY-\frac{1}{2}\bpartial\mbY\right)C_{zz}^++\frac{\partial\mY+\bpartial\mbY}{2}2\mathcal{B}_{zz}\\
  &  \delta{\mathcal{B}_{zz}}=\left(\mY\partial+\mbY\bpartial+\frac{3}{2}\partial\mY-\frac{1}{2}\bpartial\mbY\right)\mathcal{B}_{zz} \,.\\
 %   \delta\widetilde{C}_{zz}(u,z,\bz)&=\left(\mY\partial+\mbY\bpartial+\frac{3}{2}\partial\mY-\frac{1}{2}\bpartial\mbY\right)\widetilde{C}_{zz}+\frac{u}{2}(\partial\mY+\partial\mbY)\partial_u\widetilde{C}_{zz}
\eadat
\end{equation}

Hence, comparing with the log CFT doublet transformation laws \eqref{log_doublet_def}, we can see that the presence of $ \mathcal{B}_{zz} $ modifies the transformation property of the Goldstone current $C_{zz}^+$ to that of a logarithmic primary field of weights $\left(\frac{3}{2},-\frac{1}{2}\right)$ in a doublet with $\mathcal{B}_{zz} $.
On the contrary, the transformation properties for $C_{zz}^-$ remain the one of a conformal primary as in the $u\rightarrow-\infty$ asymptotic region there is no $\log u$ contribution\footnote{This can be relaxed easily to include them as well.}. 
Defining the supertranslation Goldstone current $ \mathscr{C}_{zz}$ as the linear combination
\begin{equation}
     \mathscr{C}_{zz}=\frac{C_{zz}^++C_{zz}^-}{2}\,,
    \label{C^{(0)}_{zz}_def_1}
\end{equation}
we thus have
\begin{equation}
    \mathscr{C}'_{zz}=(\partial f)^{-\frac{3}{2}}(\bpartial \bar{f})^{\frac{1}{2}}\left( \mathscr{C}_{zz}-\log|\partial f|  \mathcal{B}_{zz} \right)
    \label{mathscr{C}_{zz}_log_property}\,.
\end{equation}

\subsection{Two-point functions}\label{subsec:two_point_functions}
We will now argue that the logarithmic primary field \eqref{C^{(0)}_{zz}_def_1} can be interpreted as an IR-regulated Goldstone current. Indeed, we will show below that the following expression
\begin{equation}
\badat{2}
    \mathscr{C}_{zz}=&\frac{i\kappa}{8\pi^2}\Big[\int_0^\infty d\omega\,\omega^{2\epsilon}(a_+(\omega,z,\bz)-a_-^\dagger(\omega,z,\bz))\\
    &\quad \quad +\int\frac{d^{2+2\epsilon}y}{\pi}\frac{\bz-\by}{(z-y)^3}|z-y|^{-2\epsilon}\int_0^\infty d\omega\,\omega^{\epsilon}(a_-(\omega,y,\by)-a_+^\dagger(\omega,y,\by))\Big]\,
    \label{C_regulated_operator}
    \eadat
\end{equation}
transforms as a logarithmic primary field. In the above, we used dimensional regularization with $d=2+2\epsilon$ and one can see that \eqref{C_regulated_operator} consists of the sum of two primary operators, with conformal weights $\left(\frac{3}{2}+\epsilon,-\frac{1}{2}+\epsilon\right)$ and $\left(\frac{3}{2}+\frac{\epsilon}{2},-\frac{1}{2}+\frac{\epsilon}{2}\right)$ respectively, which turn out to be the same in the limit $\epsilon \to 0$. The primary partner of $\mathscr{C}_{zz}$ can be written as
\begin{equation}
\badat{2}\label{scr_Bzz}
     \mathscr{B}_{zz}
     &=\left[\mathcal{B}_{zz}+\int \frac{d^2y}{\pi}\frac{\bz-\by}{(z-y)^3}\mathcal{B}_{\by\by}\right]\\
     &=\frac{i\kappa}{8\pi^2}\lim_{\omega\rightarrow0}\omega\left[(a_+(\omega,z,\bz)-a_-^\dagger(\omega,z,\bz))+\int \frac{d^2y}{\pi}\frac{\bz-\by}{(z-y)^3}\,(a_-(\omega,y,\bar y)-a_+^\dagger(\omega,y,\bar y))\right]
\eadat
\end{equation}
where we used \eqref{new}.

Let us check that the transformation property of $\mathscr{C}_{zz}$, in the limit $\epsilon\rightarrow0$, does correspond to the one of a logarithmic field with $\mathscr{B}_{zz}$ as primary partner. Under a conformal transformation, the field $\mathscr{C}_{zz}$ transforms as\footnote{We drop the explicit dependence of the arguments for notation simplicity.}
\begin{equation}
\badat{2}
    \mathscr{C}'_{zz}=\frac{i\kappa}{8\pi^2}(\partial f)^{-\frac{3}{2}}(\bpartial \bar f)^{\frac{1}{2}}\bigg[
    &|\partial f|^{-2\epsilon}\int_0^\infty d\omega\,\omega^{2\epsilon}(a_+-a_-^\dagger)+\\
    &|\partial f|^{-\epsilon}\int\frac{d^{2+2\epsilon}y}{\pi}\frac{\bz-\by}{(z-y)^3}|z-y|^{-2\epsilon}\int_0^\infty d\omega\,\omega^{\epsilon}(a_--a_+^\dagger)\bigg]\,.
    \label{curly_C_Tranf_proof}
\eadat
\end{equation}
Now let us focus on the first term, and expand the expression for $\epsilon$ small,
\begin{equation}
    |\partial f|^{-2\epsilon}\int_0^\infty d\omega\,\omega^{2\epsilon}(a_+-a_-^\dagger)=\int_0^\infty d\omega\,\omega^{2\epsilon}(a_+-a_-^\dagger)-\log|\partial f|2\epsilon\int_0^\infty d\omega\,\omega^{2\epsilon}(a_+-a_-^\dagger)+o(\epsilon)\,.
\end{equation}
In the second term, we notice the appearance of a $\log|\partial f|$ term multiplied by the operator
\begin{equation}
    2\epsilon\int_0^\infty d\omega\,\omega^{2\epsilon}(a_+-a_-^\dagger).
    \label{first_int}
\end{equation}
By splitting the integral in \eqref{first_int} into two regions, using a small cut-off $\lambda$:
\begin{equation}
    2\epsilon\int_0^\lambda d\omega\,\omega^{2\epsilon}(a_+-a_-^\dagger)+ 2\epsilon\int_\lambda^\infty d\omega\,\omega^{2\epsilon}(a_+-a_-^\dagger)\,,
    \label{integral_splitting}
\end{equation}
we can expand the first integral in a series around $\omega=0$. Using the soft expansion
\begin{equation}
    (a_+(\omega,z,\bz)-a_-^\dagger(\omega,z,\bz))\simeq \frac{1}{\omega}\lim_{\xi\rightarrow0}\xi(a_+(\xi,z,\bz)-a_-^\dagger(\xi,z,\bz))+\sum_{n=0}^\infty \omega^n c_{n}(z,\bz)
\end{equation}
and assuming that this expression is convergent in a small radius around $\omega=0$, we can plug it in the first term of \eqref{integral_splitting} to get
\begin{equation}
\badat{2}
   & 2\epsilon\int_0^\lambda d\omega\,\omega^{2\epsilon-1}\lim_{\xi\rightarrow0}\xi(a_+(\xi,z,\bz)-a_-^\dagger(\xi,z,\bz))+\sum_{n=0}^\infty 2\epsilon\int_0^\lambda d\omega\,\omega^{2\epsilon+n} c_{n}(z,\bz)\\
   & =\lambda^\epsilon\lim_{\xi\rightarrow0}\xi(a_+-a_-^\dagger)+\sum_{n=0}^\infty \frac{2\epsilon}{2\epsilon+1+n}\lambda^{2\epsilon+n+1}c_{n}(z,\bz)=\lim_{\xi\rightarrow0}\xi(a_+-a_-^\dagger)+o(\epsilon).
   \eadat
\end{equation}
This implies that, as no other $1/\epsilon$ poles will come from the UV region $(\lambda,+\infty)$, \eqref{first_int} reduces to
\begin{equation}
    2\epsilon\int_0^\infty d\omega\,\omega^{2\epsilon}(a_+-a_-^\dagger)=\lim_{\xi\rightarrow0}\xi(a_+-a_-^\dagger)+o(\epsilon)\,.
\end{equation}
Notice that this is equivalent to writing
\begin{equation}
    \lim_{\Delta\rightarrow1}(\Delta-1)\int_0^\infty d\omega\,\omega^{\Delta-1}(a_+-a_-^\dagger)=\lim_{\xi\rightarrow0}\xi(a_+-a_-^\dagger)\,,
    \label{delta->1_and_omega->0}
\end{equation}
which is the statement that soft modes can be extracted as poles of Mellin-transformed operators \cite{Pate:2019mfs,Fotopoulos:2019vac}.
Using this result we can then write
\begin{equation}
    |\partial f|^{-2\epsilon}\int_0^\infty d\omega\,\omega^{2\epsilon}(a_+-a_-^\dagger)=\int_0^\infty d\omega\,\omega^{2\epsilon}(a_+-a_-^\dagger)-\log|\partial f|\lim_{\xi\rightarrow0}\xi(a_+-a_-^\dagger)+o(\epsilon)\,.
    \label{celestial_log_transform_proof}
\end{equation}
Plugging this back into \eqref{curly_C_Tranf_proof} and using a similar argument for the second term, we then have
\begin{equation}
\badat{2}
    \mathscr{C}'_{zz}
    &=(\partial f)^{-\frac{3}{2}}(\bpartial \bar f)^{\frac{1}{2}}\bigg[\mathscr{C}_{zz}-\\
    &\log|\partial f|\frac{i\kappa}{8\pi^2}\left(\lim_{\xi\rightarrow0}\xi(a_+-a_-^\dagger)+\int \frac{d^{2+2\epsilon}y}{\pi}\frac{\bz-\by}{(z-y)^3}|z-y|^{-2\epsilon}\lim_{\xi\rightarrow0}\xi(a_--a_+^\dagger)\right)+o(\epsilon)\bigg]\,.
\eadat
\end{equation}
This implies that for $\epsilon\rightarrow0$ and using \eqref{scr_Bzz}, we are left with
\begin{equation}
\badat{2}
    \mathscr{C}'_{zz}
    &=(\partial f)^{-\frac{3}{2}}(\bpartial \bar f)^{\frac{1}{2}}\bigg[\mathscr{C}_{zz}-
    \log|\partial f|\frac{i\kappa}{8\pi^2}\lim_{\xi\rightarrow0}\xi\left((a_+-a_-^\dagger)+\int \frac{d^{2}y}{\pi}\frac{\bz-\by}{(z-y)^3}(a_--a_+^\dagger)\right)\bigg]\\
    &=(\partial f)^{-\frac{3}{2}}(\bpartial \bar f)^{\frac{1}{2}}\bigg[\mathscr{C}_{zz}-\log|\partial f|\mathscr{B}_{zz}\bigg]\,,
\eadat
\end{equation}
which is the transformation property of a logarithmic primary of weights $(\frac{3}{2},-\frac{1}{2})$ partnered with $\mathscr{B}_{zz}$.
In appendix \ref{Appendix_C}, we provide an alternative definition of $\mathscr{C}_{zz}$ related to derivative operators of the form $\partial_\Delta{\mathcal{O}^{(\Delta)}}$, with $\mathcal O^{(\Delta)}$ a primary operator of conformal dimension $\Delta$. 

Let us now turn to the two-point functions of the log pair.
Computing first the $\langle\mathscr{B}_{zz}\mathscr{B}_{ww}\rangle$ two-point function, we have
\begin{equation}\label{BB}
    \badat{2}
    \langle\mathscr{B}_{zz}\mathscr{B}_{ww}\rangle
    &=-\frac{\kappa^2}{64\pi^4}\bigg\langle\lim_{\omega\rightarrow0}\omega\left[a_+-a_-^\dagger+\int \frac{d^2x}{\pi}\frac{\bz-\bx}{(z-x)^3}\left(a_--a_+^\dagger\right)\right]\\
    &\hspace{1.8cm}\times\lim_{\xi\rightarrow0}\xi\left[a_+-a_-^\dagger+\int \frac{d^2y}{\pi}\frac{\by-\bw}{(y-w)^3}\,(a_--a_+^\dagger)\right]\bigg\rangle\\
    &=-\frac{\kappa^2}{64\pi^5}\Bigg[\int d^2y\frac{\bw-\by}{(w-y)^3}\lim_{\omega,\xi\rightarrow0}\omega\xi\langle(a_--a_+^\dagger)(a_+-a_-^\dagger)\rangle\\
    &\quad \quad \quad +\int d^2x\frac{\bz-\bx}{(z-x)^3}\lim_{\omega,\xi\rightarrow0}\omega\xi\langle(a_+-a_-^\dagger)(a_--a_+^\dagger)\rangle \Bigg]\\
    &=\frac{\kappa^2}{2\pi^2}\frac{\bz-\bw}{(z-w)^3}\lim_{\omega\rightarrow0}\omega\delta(\omega)=0\,,
\eadat
\end{equation}
which is expected from the primary field in a logarithmic doublet.
We then find 
\begin{equation}
\badat{2}
    \langle\mathscr{C}_{zz}\mathscr{B}_{ww}\rangle
    &=-\frac{\kappa^2}{64\pi^4}\bigg\langle\int_0^\infty d\omega\left[\omega^{2\epsilon}\left(a_+-a_-^\dagger\right)+\int \frac{d^{2+2\epsilon}x}{\pi}\frac{\bz-\bx}{(z-x)^3}|z-x|^{-2\epsilon}\omega^{\epsilon}\left(a_--a_+^\dagger\right)\right]\times\\
    &\hspace{2.1cm}\times\lim_{\xi\rightarrow0}\xi\left[a_+-a_-^\dagger+\int \frac{d^{2+2\epsilon}y}{\pi}\frac{\bw-\by}{(w-y)^3}\,(a_--a_+^\dagger)\right]\bigg\rangle\\
    &=-\frac{\kappa^2}{64\pi^5}\Bigg[\int d^{2+2\epsilon}y\frac{\bw-\by}{(w-y)^3}\int_0^{\infty}d\omega\omega^{2\epsilon}\lim_{\xi\rightarrow0}\xi\langle(a_--a_+^\dagger)(a_+-a_-^\dagger)\rangle\\
    &\quad\quad\quad+\int d^{2+2\epsilon}x\frac{\bz-\bx}{(z-x)^3}|z-x|^{-2\epsilon}\int_0^{\infty}d\omega\omega^\epsilon\lim_{\xi\rightarrow0}\xi\langle(a_+-a_-^\dagger)(a_--a_+^\dagger)\rangle\Bigg]\\
    &=\frac{\kappa^2}{4\pi^2}\left[\frac{\bz-\bw}{(z-w)^3}+\frac{\bz-\bw}{(z-w)^3}|z-w|^{-2\epsilon}\lim_{\xi\rightarrow0}\xi^{-\epsilon}\right]\,.
\eadat
\end{equation}
Hence, using that as $\epsilon\rightarrow 0$, $\lim_{\xi\rightarrow0}\xi^{-\epsilon}=1$ we conclude
\begin{equation}\label{CB}
    \langle\mathscr{C}_{zz}\mathscr{B}_{ww}\rangle=\frac{\kappa^2}{2\pi^2}\frac{\bz-\bw}{(z-w)^3}\,.
\end{equation}
\noindent
Finally, we turn to the two-point function $\langle \mathscr{C}_{zz}\mathscr{C}_{ww}\rangle$,
\begin{equation}
\badat{2}
    \langle \mathscr{C}_{zz}\mathscr{C}_{ww}\rangle=
    &-\frac{\kappa^2}{64\pi^5}\int_0^{\infty} d\omega d\xi \,\omega^{2\epsilon}\xi^{\epsilon}\int d^2x \frac{\bz-\bx}{(z-x)^3}|z-x|^{-2\epsilon}\langle (a_--a_+^\dagger)(a_+-a_-^\dagger)\rangle\\
    &-\frac{\kappa^2}{64\pi^5}\int_0^{\infty} d\omega d\xi \,\omega^{\epsilon}\xi^{2\epsilon}\int d^2x \frac{\bw-\by}{(w-y)^3}|w-y|^{-2\epsilon}\langle (a_+-a_-^\dagger)(a_--a_+^\dagger)\rangle\\
    =&\frac{\kappa^2}{2\pi^2}\frac{\Lambda^\epsilon}{\epsilon}\frac{\bz-\bw}{(z-w)^3}|z-w|^{-2\epsilon}\\
    =&\frac{\kappa^2}{2\pi^2}\frac{\bz-\bw}{(z-w)^3}\left(\frac{1}{\epsilon}-\log|z-w|^2\right)+o(\epsilon)\,,
    \eadat
\end{equation}
where, in the last step, we have expanded for $\epsilon\rightarrow0$ and neglected the UV divergence. Combining this last equation with \eqref{BB} and \eqref{CB}, we thus recognise the logarithmic doublet structure of the form \eqref{eq:log_doublet_two_point_function} with $\tilde k_{\mathcal O}=\epsilon^{-1}$ and $k_{\mathcal O}=\frac{\kappa^2}{2\pi^2}$. As discussed there, we know that we should not expect the divergence in $\tilde k_{\mathcal O}$ to be physical as it can always be reabsorbed with a shift $\mathscr{C}_{zz}\to \mathscr{C}'_{zz}=\mathscr{C}_{zz}-\frac{\tilde k_{\mathcal O}}{2} \mathscr{B}_{zz} $. As a result, we arrive at the following IR-finite two-point functions of a logarithmic doublet of weights $(\frac{3}{2},-\frac{1}{2})$, namely
\begin{equation}
\badat{2}
    &\langle \mathscr{C}'_{zz}\mathscr{C}'_{ww}\rangle=-\frac{\kappa^2}{2\pi^2}\frac{\bz-\bw}{(z-w)^3}\log|z-w|^2\\
    &\langle \mathscr{C}'_{zz}\mathscr{B}_{ww}\rangle=\frac{\kappa^2}{2\pi^2}\frac{\bz-\bw}{(z-w)^3}\\
    &\langle \mathscr{B}_{zz}\mathscr{B}_{ww}\rangle=0\,.
\eadat
\end{equation}

\section{Discussion}
\label{sec:discussion}

\quad In this paper, we have shown that the presence of the operator \eqref{new} induces a logarithmic CFT doublet structure in the soft sector of celestial CFT. While we focused on the case of gravity, our analysis carries along for the QED case as well. The quantization of this operator was shown in \cite{Campiglia:2019wxe,AtulBhatkar:2019vcb} to source the fields at timelike infinity and to induce the quantum part of Sahoo-Sen's logarithmic soft photon theorem. It was also pointed out there that the $\log u$ quantum mode is also closely related to the asymptotic dressed operators considered in the Faddeev-Kulish construction \cite{Kulish:1970ut}. 
Recently, logarithmic corrections to the subleading soft graviton theorem were derived from the Ward identity associated with superrotation charge conservation~\cite{Agrawal:2023zea}. It is interesting to note that the derivation there did not require the inclusion of a $\log u$ operator; the fall-off conditions considered there for the gravitational shear at $u\to \infty$ were only relaxed so as to encompass gravitational tail effects (namely $\sim u^{-1}$). The object that played a central role in the derivation of \cite{Agrawal:2023zea} (see also \cite{Donnay:2022hkf,Pasterski:2022djr}) was rather the supertranslation Goldstone two-point function. The latter appeared because the superrotation charge involves an extra contribution accounting for the dressing of operators by the Goldstone boson, which is obviously related to the Faddeev-Kulish construction as well.  It would thus be interesting to clarify the connection between different derivations of logarithmic corrections to soft theorems \cite{Campiglia:2019wxe,AtulBhatkar:2019vcb,Donnay:2022hkf,Pasterski:2022djr,Agrawal:2023zea,Choi:2024ygx} based on symmetry principles. From what we have discussed, logarithmic CFT structures can be expected to play an important role once accounting for (quantum) loop corrections. This is also in line with the analyses of \cite{Bhardwaj:2022anh,Bhardwaj:2024wld} which have unveiled the presence of logarithmic CFT operators in IR-finite gluon OPEs in CCFT at one-loop.

We conclude this section by discussing below an analogy between the structure we have found with the case of a free scalar in $2d$ CFT when considering the divergent part of the correlator associated to the scalar zero-mode. This is followed by a short discussion on the appearance of logarithmic correlation functions in Coulomb gas models when the Liouville puncture operator is included in the spectrum of a gravitationally dressed CFT.

\subsection{The example of the free scalar field}

We begin this section with a recap of the treatment of the free scalar field in dimensional regularization. The action for the free scalar field in $d$-dimension is given by
\begin{equation}
    S=\frac{1}{2}\int d^dx\, \partial_\mu\Phi(x)\partial^\mu\Phi(x).
\end{equation}
The two-point correlation functions of the scalar field takes then the following form
\begin{equation} \label{twopointfree}
    \langle\Phi(x)\Phi(y)\rangle=\pi^{\frac{d}{2}-1}\Gamma\left(\frac{d-2}{2}\right)\frac{1}{|x-y|^{d-2}}.
\end{equation}
In dimensional regularization near $2d$, we take $d=2+2\epsilon$ and expanding \eqref{twopointfree} for $\epsilon\rightarrow0$ we get
\begin{equation}
    \langle\Phi(z,\bz)\Phi(w,\bw)\rangle=\frac{1}{\epsilon}-\log|z-w|^2\mu^2+o(\epsilon)
    \label{eq:free_scalar_two_point}
\end{equation}
where $\mu$ is the mass scale which enforces the argument to be dimensionless. Notice that, in this setup, there is a divergent part followed by the usual logarithmic two-point function. In standard expressions (see \cite{DiFrancesco1997nk}), the divergent term is typically left inside the logarithm or not written explicitly.
In order to extract the divergent part of the correlator, we can define the field  $\varphi=\epsilon\Phi$ such that the two-point correlator becomes
\begin{equation}
\begin{alignedat}{2}
    \langle\Phi(z, \bz)\varphi(w,\bw)\rangle&=1-\epsilon\log|z-w|^2\mu^2\xrightarrow{\epsilon\rightarrow0}1\\
    \langle\varphi(z,\bz)\varphi(w,\bw)\rangle&=\epsilon-\epsilon^2\log|z-w|^2\mu^2\xrightarrow{\epsilon\rightarrow0}0.
\end{alignedat}
\end{equation}
From the expressions above, it is apparent that $\varphi$ extracts the divergent part induced by the presence of the zero mode of $\Phi$. To make it even more explicit, it is possible to expand $\Phi$ in modes (see e.g. \cite{DiFrancesco1997nk, Maccaferri:2023wrg})
\begin{equation}
    \Phi(z,\bz)=\chi-i\phi_0\log{|z|^2}+i\sum_{n\neq0}\left(\frac{1}{n}\phi_nz^{-n}+\frac{1}{n}\bar{\phi}_n\bz^{-n}\right)\,,
\end{equation}
with the commutation rules $[\chi,\phi_0]=i$ and $[\phi_n,\phi_m]=[\bar{\phi}_n,\bar{\phi}_m]=n\delta_{n+m,0},\,[\phi_n,\bar{\phi}_m]=0$\,.
To enforce \eqref{eq:free_scalar_two_point} the zero mode $\chi$ has to satisfy 
\begin{equation}
    \langle\chi\chi\rangle=\frac{1}{\epsilon}\,,
\end{equation}
while all the other modes give a finite contribution acting on the vacuum. We can conclude then that in the limit $\epsilon\rightarrow0$ the operator $\varphi$ extracts the contribution of the zero mode as
\begin{equation}
    \varphi(z,\bz)=\epsilon\Phi(z,\bz)=\epsilon\chi+\epsilon\left(-i\phi_0\log{|z|^2}+i\sum_{n\neq0}\left(\frac{1}{n}\phi_nz^{-n}+\frac{1}{n}\bar{\phi}_n\bz^{-n}\right)\right)\xrightarrow[]{\epsilon\rightarrow0}\epsilon\chi\,,
    \label{varphi_as_zero_mode}
\end{equation}
and all the other operators evaluate to zero in this limit.

The example of the free scalar field is suggestive of the behavior of the Goldstone mode. To make it manifest, we define the normalised mode
\begin{equation}
    \widehat{C}_\epsilon=\sqrt{\epsilon}C_\epsilon
\end{equation}
such that at leading order in the $\epsilon$ the correlator is 
\begin{equation}
\badat{3}
    \langle \widehat{C}_\epsilon(z,\bz) \widehat{C}_\epsilon(w,\bw)\rangle
    &=\frac{\kappa^2}{16\pi^2}|z-w|^2\left(\frac{1}{\epsilon}+\log(|z-w|^2\mu^2)\right)\,.
\eadat
\end{equation}
This structure resembles the two-point function of the free scalar. Following the same reasoning as in \eqref{varphi_as_zero_mode}, we can then relate the following operator $\widehat{B}_\epsilon=\epsilon \widehat{C}_\epsilon$ with the $\widehat{C}_\epsilon$ zero mode.
Moreover, we can use \eqref{delta->1_and_omega->0} to rewrite  $\widehat{B}_\epsilon$ in the limit $\epsilon\rightarrow0$ as follows
\begin{equation}
    \widehat{B}_\epsilon(z,\bz)=\frac{\sqrt{\epsilon}}{2}\int \frac{d^{2+2\epsilon}w}{2\pi}\left[\frac{z-w}{\bz-\bw}\mathcal{B}_{ww}+\frac{z-w}{\bz-\bw}\mathcal{B}_{ww}\right]\,,
\end{equation}
then making it clear that the mode $\mathcal{B}_{zz}$ is needed in the construction of the zero mode $\widehat{B}_\epsilon$ in this regularization scheme.

\subsection{Liouville theory in CCFT}
\quad As discussed in \cite{Compere:2016jwb,Donnay:2021wrk,Campiglia:2021bap,Campiglia:2020qvc,Compere:2020lrt} (see also eq.~\eqref{eqasympt}), the inclusion of superrotations in the soft gravitational phase space involves the vacuum news tensor $N_{zz}^{\text{vac}}$, a $(2,0)$ mode which has the transformation properties of a $2d$ stress-tensor,
\begin{equation}
\delta_{\mY}N_{zz}^{\text{vac}}=\left(\mY\partial+2\partial\mY\right)N_{zz}^{\text{vac}}-\partial^3\mY\,.
\end{equation}
It is related to the so-called `superboost scalar field' $\Phi$~\cite{Compere:2016jwb}, which satisfies
\begin{equation}
    D_AD^A \Phi(z,\bz)=\mathring{R} \quad \rightarrow \quad \Phi(z,\bz)=\varphi(z)+\bar\varphi(\bz)-\log\Omega(z,\bz),
\end{equation}
where $D_A$ is the covariant derivative with respect to the $2d$ metric $\mathring{q}_{AB}dx^Adx^B=2\Omega(z,\bz)\, dz d\bz$ of Ricci tensor $\mathring{R}$.
The vacuum news tensor can be expressed in terms of the holomorphic part of $\Phi$ as
\begin{equation}
    N^{\text{vac}}_{zz}=\frac{1}{2}(D_z\varphi)^2-D_z^2\varphi\,.
\end{equation}
The superboost field $\varphi(z)$ was used in \cite{Fiorucci:2023lpb} to construct a composite operator which was argued to form a logarithmic pair with the celestial stress tensor.
It has also been noted in \cite{Compere:2016jwb,Compere:2018ylh}  that $N^{vac}_{zz}$ is precisely proportional to the stress tensor of a Euclidean Liouville theory defined by the action 
\begin{equation}\label{S_Liouville}
    S=\frac{\gamma^2}{4\pi}\int d^2x \sqrt{\mathring{q}}\left[\frac{1}{2}D_A\Phi D^A\Phi+\mathring{R} \Phi\right],
\end{equation}
where we introduced a parameter $\gamma$.
Notice that since the Liouville cosmological constant $\Lambda$ is zero in \eqref{S_Liouville}, the action reduces to that of a Coulomb gas.  
In particular, we can make a field redefinition $\Phi=\phi/\gamma$ to bring the action in the form
\begin{equation}
     S=\frac{1}{4\pi}\int d^2x \sqrt{\mathring{q}}\left[\frac{1}{2}D_A\phi D^A\phi+\gamma\mathring{R}\phi\right]\,.
\end{equation}
Following the notation of \cite{DiFrancesco1997nk}, we see that this is a Coulomb gas with central charge
\begin{equation}
    c=1+12\gamma^2=1-24\alpha_0^2\,,
\end{equation}
where we introduced $\alpha_0$ such that $\gamma=i\sqrt{2}\alpha_0$ to make contact with standard notation.

We want to analyze the properties of the operator content of this theory, which is usually represented by vertex operators
\begin{equation}
    V_{\alpha}=:e^{i\sqrt{2}\alpha\phi}:
    \label{vertex_operator}
\end{equation}
of conformal weight
\begin{equation}\label{h_alpha}
    h_\alpha=\alpha(\alpha-2\alpha_0)=\alpha(\alpha+i\sqrt{2}\gamma)\,.
\end{equation}
For the moment, we will be agnostic about the value of $\gamma$ but we remark that, if $\gamma$ is purely imaginary, then vertex operators with real conformal weights will be those with $\alpha\in\mathbb{R}$, while for $\gamma\in\mathbb{R}$ we will have to consider purely imaginary $\alpha$. In this case, vertex operators turn into those usually considered in Liouville theory, $V_{\lambda}(z)=:e^{\lambda\phi}:(z)$ with $\lambda\in\mathbb{R}$.

We can now discuss the appearance of logarithmic operators in Coulomb gas models, as studied in \cite{Kogan:1997fd}. We first recall that, in this theory, only correlators which satisfy  the neutrality condition $\sum \alpha_i=2\alpha_0$ are non-vanishing; in particular,  the only non-zero two-point functions will be those of the form $\langle V_{\alpha} V_{2\alpha_0-\alpha}\rangle$~\cite{DiFrancesco1997nk}. From \eqref{h_alpha}, we see that the operators $V_{\alpha}$ and $V_{2\alpha_0-\alpha}$ have the same conformal dimension and when $\alpha=\alpha_0$, the two operators degenerate to the same one. It turns out that, for this value of $\alpha$, there exists a second operator, 
\begin{equation}\label{puncture}
    V_P=:\phi\, e^{i\sqrt{2}\alpha_0\phi}:=\frac{1}{i\sqrt{2}}\frac{d V_\alpha}{d\alpha}\bigg|_{\alpha=\alpha_0}\,,
\end{equation}
which is also a \emph{primary} field of the same dimension than $V_{\alpha_0}$, namely $h_{\alpha_0}=-\alpha_0^2$. In Liouville theory, where $\alpha_0$ is imaginary, $V_P$ is known as the `puncture operator'~\cite{Zamolodchikov:1995aa}.
Notice that $\frac{d V_\alpha}{d\alpha}$ will be a primary field \emph{only} if $\alpha=\alpha_0$, while for generic values of $\alpha$ it will turn into a logarithmic primary. This is easy to see from the relation between $h_\alpha$ and $\alpha$ as we can rewrite
\begin{equation}
    \frac{d V_\alpha}{d\alpha}=2(\alpha-\alpha_0)\frac{d V_\alpha}{dh}=4(\alpha-\alpha_0)\frac{d V_\alpha}{d\Delta}\,,
\end{equation}
which shows that for $\alpha\neq\alpha_0$, $\frac{d V_\alpha}{d\alpha}$ is related to operators of the form $\partial_\Delta \mathcal{O}^{(\Delta)}$ (with $\mathcal{O}^{(\Delta)}$ a primary) which are known to be logarithmic\footnote{This is easy to check from the transformation properties of primary fields
\begin{equation}
\badat{2}\nonumber
    \partial_\Delta{\mathcal{O}^{(\Delta)}}'(z',\bz')
    &=\left(\frac{\partial f}{\bpartial\bar{f}}\right)^{-J}\partial_\Delta\left(|\partial f|^{-\Delta}\mathcal{O}^{(\Delta)}(z,\bz)\right)\\
    &=(\partial f)^{-h}(\bpartial \bar{f})^{-\bh}\left[\partial_\Delta \mathcal{O}^{(\Delta)}+\log|\partial f|\mathcal{O}^{(\Delta)}\right]\,.
\eadat
\end{equation}} \cite{Hogervorst:2016itc,Flohr:2001zs,Nivesvivat:2020gdj}. 
While the puncture operator \eqref{puncture} is an ordinary primary field, its inclusion in the spectrum will give rise to logarithmic correlation functions \cite{Kogan:1997fd,Giribet:2001qq,Giribet:2004qe}. For this reason, it is sometimes referred to as a `pre-logarithmic' operator, namely a primary operator whose OPEs with other primaries will contain logarithmic operators.
An interesting case in this context is the one of a gravitationally dressed CFT where $\alpha_0^2<1$ and the Coulomb gas is used to couple another a CFT with $2d$ gravity \cite{Bilal:1994nx}. If we denote by $\Phi_h$ the operators of the CFT coupled to gravity, the dressed operators will take the form
\begin{equation}
    \mathcal{O}(z)=:V_\alpha\Phi_h:
\end{equation}
with conformal weight equal to $h_\alpha+h=1$. This means that, if $h=1+\alpha_0^2$, then $h_\alpha=-\alpha_0^2$ and one thus needs to also include the puncture operator to dress the fields, which will give rise to a logarithmic structure~\cite{Bilal:1994nx,Kogan:1997fd}. 
If the superrotation Liouville field can be interpreted as some gravitational dressing (see \cite{Choi:2019rlz,Pasterski:2021dqe,Freidel:2022skz} for subleading conformally soft dressings),
then the above discussion would allow to understand the appearance of logarithmic operators in CCFT as a result of the inclusion in the spectrum of the Liouville puncture operator. It would also be interesting to explore possible connections with the recent works highlighting the use of Liouville theory in celestial holography~\cite{Stieberger:2022zyk,Stieberger:2023fju,Giribet:2024vnk,Melton:2024gyu}. 

 \section*{Acknowledgments}
 We thank Miguel Campiglia, Lorenzo Di Pietro, Laurent Freidel, Gaston Giribet, Yannick Herfray, Alok Laddha, Prahar Mitra, Sruthi Narayanan, Miguel Paulos, Andrea Puhm, Ana-Maria Raclariu and Marco Serone for useful discussions and comments.
 The work of A.B. is partially funded by the Knut and Alice Wallenberg Foundation grant KAW 2021.0170 and Olle Engkvists Stiftelse grant 2180108. L.D. are B.V. are supported by the European Research Council (ERC) Project 101076737 -- CeleBH. Views and opinions expressed are however those of the authors only and do not necessarily reflect those of the European Union or the European Research Council. Neither the European Union nor the granting authority can be held responsible for them.
 L.D. and B.V. are also partially supported by INFN Iniziativa Specifica ST\&FI.  L.D.'s research was also supported in part by the Simons Foundation through the Simons Foundation Emmy Noether Fellows Program at Perimeter Institute. L.D. and B.V. are grateful for the hospitality of Perimeter Institute where part of this work was done. Research at Perimeter Institute is supported in part by the Government of Canada through the Department of Innovation, Science and Economic Development and by the Province of Ontario through the Ministry of Colleges and Universities. Part of this project was performed during the Programme ``Carrollian Physics and Holography'' at the Erwin-Schrödinger International Institute for Mathematics and Physics in April 2024.

\appendix

\section{More on the log-shadow}
\label{sec:more_on_the_log_shadow}

In this appendix we would like to prove that $\widetilde{\Psi}$ transforms as a logarithmic primary of weights $(1-h,1-\bh)$. Under a generic $SL(2,\mathbb{C})$ transformation:
\begin{equation}
\badat{2}
  &  z'=f(z)=\frac{a z+b}{c z+d},\quad \bz'=\bar f(\bz)=\frac{\ba \bz+\bb}{\bc \bz+\bd}\\
    &\partial f(z)=\frac{1}{(cz+d)^2},\quad \bar \partial \bar f(\bz)=\frac{1}{(\bc\bz+\bd)^2}\,
\eadat\end{equation}
we have
\begin{equation}
    \widetilde{\Psi}'(z',\bz')=-K_{h,\bh}\int d^2w \frac{\Psi'(w,\bw)+\mathrm{log}|z'-w|^2\Phi(w,\bw)}{(z'-w)^{2-2h}(\bz'-\bw)^{2-2\bh}}\,.
    \label{eq:shadow_log_prove_1}
\end{equation}
Changing the integration variable to $x'=w=f(x),\bx'=\bw=\bar f(\bx)$, we can use the following property:
\begin{equation}
    \left|\det\frac{\partial w}{\partial x}\right|=\frac{1}{(cx+d)^2(\bc\bx+\bd)^2};\quad z'-x'=f(z)-f(x)=\frac{z-x}{(cz+d)(cx+d)}
\end{equation}
to rewrite \eqref{eq:shadow_log_prove_1} as:
\begin{equation}
\badat{2}
       \widetilde{\Psi}(z',\bz')=-K_{h,\bh}\int d^2x\frac{(cx+d)^{-2h}(\bc\bx+\bd)^{-2\bh}}{(z-x)^{2-2h}(\bz-\bx)^{2-2\bh}}\left[\Psi'(x',\bx')+\mathrm{log}\frac{|z-x|^2}{|cz+d|^2|cx+d|^2}\Phi'(x',\bx')\right]
\eadat\end{equation}
We can then substitute the transformed fields with the identities \eqref{eq:log_doublet_tranformations} and rewrite:
\begin{equation}
\badat{2}
    \widetilde{\Psi}'(z',\bz')\,=\,&-K_{h,\bh}(\partial f)^{h-1}(\bar\partial \bar f)^{\bh-1}\int d^2x\frac{1}{(z-x)^{2-2h}(\bz-\bx)^{2-2\bh}}\\
    \times&\left[\Psi(x,\bx)-\log|\partial f(x)|\Phi(x,\bx)+\mathrm{log}\left(|z-x|^2|\partial f(z)||\partial f(x)|\right)\Phi(x,\bx)\right]\\
    =&-K_{h,\bh} (\partial f)^{h-1}(\bar\partial \bar f)^{\bh-1}\int d^2x\frac{\Psi(x,\bx)+\mathrm{log}\left(|z-x|^2|\partial f(z)|\right)\Phi(x,\bx)}{(z-x)^{2-2h}(\bz-\bx)^{2-2\bh}}\\
    =&(\partial f)^{h-1}(\bar\partial \bar f)^{\bh-1}\left[\widetilde{\Psi}(z,\bz)-\mathrm{log}|\partial f(z)|\widetilde{\Phi}(z,\bz)\right]\,,
\eadat
\end{equation}
which is exactly the transformation property of a $(1-h,1-\bh)$ log primary.\\
The shadow logarithmic doublet $\widetilde{\mathcal O}_a=(\widetilde \Psi, \widetilde \Phi)$ thus transforms as a log CFT doublet of weights $(1-h,1-\bh)$.\\
We can also be interested into computing the square of the log-shadow to verify if it also squares to 1. We need to compute:
\begin{gather}
    \widetilde{\widetilde{\Psi}}(z,\bz)
    =K_{1-h,1-\bh}K_{h,\bh}\int d^2w d^2x\frac{\Psi(x,\bx)}{(z-w)^{2h}(\bz-\bw)^{2\bh}(w-x)^{2-2h}(\bw-\bx)^{2-2\bh}}\,+\label{eq:log_square_proof_line_1}\\
    +\,K_{1-h,1-\bh}K_{h,\bh}\int d^2w d^2x\frac{\Phi(x,\bx)}{(z-w)^{2h}(\bz-\bw)^{2\bh}(w-x)^{2-2h}(\bw-\bx)^{2-2\bh}}\log\frac{|w-x|^2}{|z-w|^2}.\nonumber
\end{gather}
Because $J=h-\bh\in\mathbb{Z}/2$ the integral in \eqref{eq:log_square_proof_line_1} can be computed using the following formula found in \cite{Dolan:2011dv}:
\begin{equation}
    I_1=\int d^2w \frac{1}{(z-w)^{2h}(\bz-\bw)^{2\bh}(w-x)^{2-2h}(\bw-\bx)^{2-2\bh}}=(-1)^{-4\bar h}\pi^2\frac{\Gamma(1-2h)\Gamma(2h-1)}{\Gamma(2\bh)\Gamma(2-2\bh)}\delta^2(z-x)
\end{equation}
so that using the definition of $K_{h,\bar h}$ \eqref{eq:log_square_proof_line_1} turns out be equal to $(-1)^{-4\bar h}\Psi$. If we assume to work only with semi-integers conformal weights then it follows that the usual shadow squares to 1.\\
To compute \eqref{eq:log_square_proof_line_1} we notice that:
\begin{equation}
    \frac{1}{2}(\partial_h+\partial_\bh)I_1=\frac{\partial I_1}{\partial \Delta}=\int d^2w \frac{1}{(z-w)^{2h}(\bz-\bw)^{2\bh}(w-x)^{2-2h}(\bw-\bx)^{2-2\bh}}\log\frac{|w-x|^2}{|z-w|^2}=I_2
\end{equation}
where $\Delta=h+\bh$. This implies:
\begin{align}    
    I_2
    &=\pi^2\delta^2(z-x)(-1)^{-4\bh}\frac{\Gamma(1-2h)\Gamma(2h-1)}{\Gamma(2\bh)\Gamma(2-2\bh)}\times\\
    \times&\left(\frac{1}{1-2h}+\frac{1}{1-2\bh}-2\pi i-\pi \cot{2\pi h}+\pi \cot{2\pi \bh}\right).\nonumber
\end{align} 
Because $h$ and $\bh$ always differ by a semi-integer the cotangent part of this expression can be dropped and we get:
\begin{equation}
    I_2=\pi^2\delta^2(z-x)(-1)^{-4\bh}\frac{\Gamma(1-2h)\Gamma(2h-1)}{\Gamma(2\bh)\Gamma(2-2\bh)}\left(\frac{1}{1-2h}+\frac{1}{1-2\bh}-2\pi i\right)
\end{equation}
making the expression for the squared log-shadow:
\begin{equation}
    \widetilde{\widetilde{\Psi}}(z,\bz)=(-1)^{-4\bar h}\left[\Psi(z,\bz)+\left(\frac{1}{1-2h}+\frac{1}{1-2\bh}-2\pi i\right)\Phi(z,\bz)\right].
\end{equation}
We can clearly see that it does not square to the identity in the general case, but it shifts the log primary with the partner primary. However we can see that the inverse shadow for the doublet is well defined, as:
\begin{align}
    &S_{\log}^{-1}\left[\Psi\right](z,\bz)=(-1)^{4\bar h}\left[\widetilde{\Psi}(z,\bz)-\left(\frac{1}{1-2h}+\frac{1}{1-2\bh}-2\pi i\right)\widetilde\Phi(z,\bz)\right]\\
    &S_{\log}^{-1}\left[\Phi\right](z,\bz)=(-1)^{4\bar h}\widetilde\Phi(z,\bar z)\,.
\end{align}

\section{Regulated integral computation}
\label{sec:Regulated_integral_computation}

In this appendix, we compute explicitly the integral \eqref{eq:I_Epsilon} in dimensional regularization $d=2+2\epsilon$. At first, let us rewrite the integral using Feynman parameters as 
\begin{equation}
    I_\epsilon=\mu_0^{2\epsilon}\int d^{d}z\frac{(z-z_1)^2(\bz-\bz_2)^2}{|z-z_1|^2|z-z_2|^2}=\mu_0^{2\epsilon}\int_0^1 du\int d^{d}z\frac{(z-z_1)^2(\bz-\bz_2)^2}{(u|z-z_1|^2+(1-u)|z-z_2|^2)^2}\,.
\end{equation}
Focusing on the $z$ integral, we re-parameterize it with the following change of variables:
\begin{equation}
    z=x+uz_1+(1-u)z_2,\quad \bz=\bx+u\bz_1+(1-u)\bz_2\,\quad z_{12}=z_1-z_2\,.
\end{equation}
This leaves us with the expression
\begin{equation}
    \int d^{2+2\epsilon}x\,\frac{|x|^4+u^2(1-u)^2|z_{12}|^2-4u(1-u)|x|^2|z_{12}|^2}{(|x|^2+u(1-u)|z_{12}|^2)^2}\,,
\end{equation}
where terms linear in $x, \bx$ have been dropped due to the parity properties of the integral.\\
The expression depends on $|x|^2$ so we can factorize the angular component and get:
\begin{equation}
    J_\epsilon=\frac{2\pi^{1+\epsilon}}{\Gamma(1+\epsilon)}\int_0^\infty dr\, r^{1+2\epsilon}\frac{r^4+R^4-4r^2R^2}{(r^2+R^2)^2}\,,
\end{equation}
where we have defined $R^2=u(1-u)|z_{12}|^2$.\\
$J_\epsilon$ is linearly divergent at infinity, as we can highlight by splitting it as:
\begin{equation}
    J_\epsilon=\frac{2\pi^{1+\epsilon}}{\Gamma(1+\epsilon)}\left[\int_0^\infty dr\, r^{1+2\epsilon}-6\int_0^\infty dr\,\frac{r^{3+2\epsilon}R^2}{(r^2+R^2)^2}\right]\,,
\end{equation}
where the second term is finite for $\epsilon<0$. As the first term is a divergent dimensionful term that does not contain any scale, in dim-reg it can be directly set to zero. Any possible finite ambiguity will be taken into account by changing $\mu_0$.\\
This leaves us with:
\begin{equation}
    J_\epsilon=-\frac{12\pi^{1+\epsilon}}{\Gamma(1+\epsilon)}\int_0^\infty dr\,\frac{r^{3+2\epsilon}R^2}{(r^2+R^2)^2}=\frac{6\pi^{2+\epsilon}(1+\epsilon)}{\Gamma(1+\epsilon)\sin\pi\epsilon}R^{2+2\epsilon}\,.
\end{equation}
$I_\epsilon$ can then be easily obtained by integrating over $u$, which gives 
\begin{equation}
    I_\epsilon=\frac{6\pi^{2+\epsilon}(1+\epsilon)\Gamma(2+\epsilon)^2}{\Gamma(1+\epsilon)\Gamma(4+2\epsilon)\sin\pi\epsilon}|z_{12}|^{2+2\epsilon}\mu_0^{2\epsilon}\,.
\end{equation}
Notice that this expression as a function of $\epsilon$ can be analytically continued also in the region $\epsilon>0$.  
If we now consider the expansion in small $\epsilon$ we get: 
\begin{equation}
\badat{2}
   I_\epsilon
   &=\pi|z_{12}|^2\left(\frac{1}{\epsilon}-\frac{2}{3}+\gamma_E+\log\pi+\log(|z_{12}|^2\mu_0^2)\right)+\dots\\
   &=\pi|z_{12}|^2\left(\frac{1}{\epsilon}+\log(|z_{12}|^2\mu^2)\right)+\dots
\eadat
\end{equation}
with $\mu^2=\pi e^{\gamma_E-\frac{2}{3}}\mu_0^2$.

\section{Relation between $\mathscr{C}_{zz}$ and $\partial_\Delta{\mathcal{O}^{(\Delta)}}$} 
\label{Appendix_C}

In section \ref{transformation_properties}, we have proved that $\mathscr{C}_{zz}$ behaves like a logarithmic primary field. The aim of this appendix is to relate the latter to logarithmic fields of the form  $\partial_\Delta \mathcal{O}^{(\Delta)}$. Such derivative operators were already considered in celestial CFT in the case of loop corrected gluon OPEs \cite{Bhardwaj:2022anh,Bhardwaj:2024wld}. The presence of the $\mathscr{B}_{zz}$ operator may give rise to their presence also for graviton operators.\\
To connect with the $\mathscr{C}_{zz}$ operator we define:
\begin{equation}
    A_\pm=\frac{i\kappa}{8\pi^2}\lim_{\Delta\rightarrow 1}\partial_\Delta\left[(\Delta-1)\left(a^{(2)}_{\Delta,\pm}-a^{(2),\dagger}_{\Delta,\pm}\right)\right]=\frac{i\kappa}{8\pi^2}\lim_{\delta\rightarrow 0}\partial_\delta\left[\delta\left(a^{(2)}_{1+\delta,\pm}-a^{(2),\dagger}_{1+\delta,\pm}\right)\right]
\end{equation}
where $a^{(2)}_{1+\delta,+},a^{(2),\dagger}_{1+\delta,+} $ are defined in \eqref{eq::Mellin_ladders}. Notice that from equation \eqref{delta->1_and_omega->0} we know that $(\Delta-1)\left(a^{(2)}_{\Delta,\pm}-a^{(2),\dagger}_{\Delta,\pm}\right)$ is regular in the limit $\Delta\rightarrow1$ so that $A_+$ is finite. We are interested at first in the transformation properties of $A_\pm$ under conformal transformations. Using the transformation properties of the Mellin ladder operators we can write
\begin{equation}
\badat{1}
    A'(z',\bz')_\pm
    &=\frac{i\kappa}{8\pi^2}(\partial f)^{-\frac{3}{2}}(\bpartial \bar{f})^{\frac{1}{2}}\lim_{\delta\rightarrow 1}\partial_\delta\left[|\partial f|^{-\delta}\delta\left(a^{(2)}_{1+\delta,\pm}-a^{(2),\dagger}_{1+\delta,\pm}\right)\right]\\
    &=\frac{i\kappa}{8\pi^2}(\partial f)^{-\frac{3}{2}}(\bpartial \bar{f})^{\frac{1}{2}}\lim_{\delta\rightarrow 1}\left[|\partial f|^{-\delta}\partial_\delta\left(\delta\left(a^{(2)}_{1+\delta,\pm}-a^{(2),\dagger}_{1+\delta,\pm}\right)\right)-\log|\partial f|\delta\left(a^{(2)}_{1+\delta,\pm}-a^{(2),\dagger}_{1+\delta,\pm}\right)\right]\\
    &=(\partial f)^{-\frac{3}{2}}(\bpartial \bar{f})^{\frac{1}{2}}\left[A_\pm-\log|\partial f|\frac{i\kappa}{8\pi^2}\lim_{\delta\rightarrow0}\delta\left(a^{(2)}_{1+\delta,\pm}-a^{(2),\dagger}_{1+\delta,\pm}\right)\right]\\
    &=(\partial f)^{-\frac{3}{2}}(\bpartial \bar{f})^{\frac{1}{2}}\left[A_\pm-\log|\partial f|B_\pm\right]\,,
    \nonumber
\eadat
\end{equation}
where as $A_+$ is finite in the $\delta\rightarrow0$ limit we get no other contribution from $|\partial f|^{-\delta}$. Then we see that $A_+$ is a logarithmic field in a doublet with the primary
\begin{equation}
    B_+=\frac{i\kappa}{8\pi^2}\lim_{\delta\rightarrow0}\delta\left(a^{(2)}_{1+\delta,\pm}-a^{(2),\dagger}_{1+\delta,\pm}\right)\,.
\end{equation}
However from \eqref{delta->1_and_omega->0} we also know that
\begin{equation}
    B_+=\frac{i\kappa}{8\pi^2}\lim_{\delta\rightarrow0}\delta\left(a^{(2)}_{1+\delta,\pm}-a^{(2),\dagger}_{1+\delta,\pm}\right)=\frac{i\kappa}{8\pi^2}\lim_{\omega\rightarrow0}\omega(a_+-a_-^\dagger)=\mathcal{B}_{zz}\,,
\end{equation}
which proves that the derivative operator $A_+$ is in a logarithmic doublet with $\mathcal{B}_{zz}$. We can then relate $A_+$ with $C^+_{zz}$. Then taking into consideration also $A_-$ and the log-shadow transform we can also suggest an alternative definition for $\mathscr{C}_{zz}$:
\begin{equation}
    \mathscr{C}_{zz}\sim \left[A_+-\int \frac{d^2y}{\pi}\frac{\bz-\by}{(z-y)^3}\left(A_-+\log|z-y|^2 \mathcal{B}_{\by\by}\right)\right]\,.
    \label{mathscrC_zz_alternative}
\end{equation}
Using the properties of the log shadow it is easy to see that this operator will also be logarithmic and in a doublet with $\mathscr{B}_{zz}$. Notice that if $\mathscr{B}_{zz}=0$ the operator defined in \eqref{mathscrC_zz_alternative} is also vanishing, which underlines the fact that the logarithmic structure is present only if $\mathscr{B}_{zz}$ is non vanishing.\\
This discussion allowed us to bridge $\partial_\Delta \mathcal{O}^{(\Delta)}$ with $\mathscr{C}_{zz}$. We now want to make some additional considerations on the $A_+$ operator. To do so we will at first massage its expression by explicitly writing the definition of the derivative:
\begin{equation}
     A_+=\lim_{\delta\rightarrow 0}\lim_{\epsilon\rightarrow 0}\frac{(\delta+\epsilon)\left(a^{(2)}_{1+\delta+\epsilon,\pm}-a^{(2),\dagger}_{1+\delta+\epsilon,\pm}\right)-\delta \left(a^{(2)}_{1+\delta,\pm}-a^{(2),\dagger}_{1+\delta,\pm}\right)}{\epsilon}.
\end{equation}
Under the assumption that the limits are well defined, we can exchange their ordering to find
\begin{equation}
    A_+=\frac{i\kappa}{8\pi^2}\lim_{\epsilon\rightarrow 0}\left[\left(a^{(2)}_{1+\epsilon,\pm}-a^{(2),\dagger}_{1+\epsilon,\pm}\right)+\frac{1}{\epsilon}\lim_{\delta\rightarrow 0}\delta\left(a^{(2)}_{1+\epsilon+\delta,\pm}-a^{(2),\dagger}_{1+\epsilon+\delta,\pm}\right)-\frac{1}{\epsilon}\lim_{\delta\rightarrow0}\delta\left(a^{(2)}_{1+\delta,\pm}-a^{(2),\dagger}\right)\right]\,.
    \nonumber
\end{equation}
For $\epsilon$ finite, the second term vanishes for $\delta\rightarrow 0$, while the last term is identical to $\mathcal{B}_{zz}$ and we are left with
\begin{equation}
\badat{2}
    A_+
    &=\lim_{\epsilon\rightarrow 0}\left[\frac{i\kappa}{8\pi^2}\left(a^{(2)}_{1+\epsilon,\pm}-a^{(2),\dagger}_{1+\epsilon,\pm}\right)-\frac{1}{\epsilon}\mathcal{B}_{zz}\right]\\
    &=\lim_{\epsilon\rightarrow0}\frac{i\kappa}{8\pi^2}\left[\int_0^\infty \omega^\epsilon(a_+-a_-^\dagger)-\frac{1}{\epsilon}\lim_{\omega\rightarrow 0}\omega(a_+-a_-^\dagger)\right]\,.
\eadat
\end{equation}
This expression shows that  $A_+$ is basically a regulated version of the small $\epsilon$ limit of:
\begin{equation}
    \int_0^\infty \omega^\epsilon(a_+-a_-^\dagger)\,.
    \label{temp}
\end{equation}
In fact,  if \eqref{temp} gives a divergent result in connected $n$-point functions due to the presence of the soft pole, $A_+$ is the same operator with the additional term $1/\epsilon\, \mathcal{B}_{zz}$ that removes the soft pole giving a finite result. Notice that regulated operators of the form \eqref{temp} are precisely those used to write the dim-reg expression for $\mathscr{C}_{zz}$ \eqref{C_regulated_operator}, which tightens the relation between $\mathscr{C}_{zz}$ and $A_+$.

\bibliographystyle{style}
\bibliography{references}

\end{document}